\documentclass[11pt,a4paper]{article}
\usepackage[pdftex]{graphics}
\usepackage{jheppub}
\usepackage{float}
\usepackage{amsmath,amssymb,amsfonts}
\usepackage{caption,subcaption}
\usepackage{comment}

\newcommand{\eq}{\begin{equation}}
\newcommand{\eqe}{\end{equation}}
\newcommand{\eqa}{\begin{eqnarray}}
\newcommand{\eqae}{\end{eqnarray}}

\newbox\charbox
\newbox\slabox
\def\s#1{{      
        \setbox\charbox=\hbox{$#1$}
        \setbox\slabox=\hbox{$/$}
        \dimen\charbox=\ht\slabox
        \advance\dimen\charbox by -\dp\slabox
        \advance\dimen\charbox by -\ht\charbox
        \advance\dimen\charbox by \dp\charbox
        \divide\dimen\charbox by 2
        \raise-\dimen\charbox\hbox to \wd\charbox{\hss/\hss}
        \llap{$#1$}
}}

\preprint{}

\title{Bootstrapping string theory EFT}
\author[1]{Li-Yuan Chiang}
\author[2,3]{Yu-tin Huang}
\author[4]{He-Chen Weng}
\affiliation[1]{Department of Physics, Yale University, New Haven, CT 06520, USA}
\affiliation[2]{Department of Physics and Center for Theoretical Physics, National Taiwan University, Taipei 10617, Taiwan}
\affiliation[3]{Physics Division, National Center for Theoretical Sciences, Taipei 10617, Taiwan}
\affiliation[4]{Department of Physics, Brown University, Providence, RI 02912, USA}

\emailAdd{yutinyt@gmail.com}
\emailAdd{he-chen\_weng@brown.edu}
\emailAdd{li-yuan.chiang@yale.edu}
\abstract{We study the space of open string effective field theories by combining the constraint of unitarity and monodromy relations for the four-point amplitude. The latter is a reflection of an underlying disk correlator with singularities at the boundary. By assuming maximal susy the resulting bootstrap isolates Wilson coefficients to at least $10^{-4}$ of the Type-I superstring. Furthermore, utilizing our geometric approach, we obtain the critical dimension of 10 from the low energy coefficients alone. Remarkably, relaxing SUSY but requiring the massless states to carry four-dimensional helicities, the Wilson coefficients are again constrained to superstring values within $10^{-4}$. Thus we conclude that type-I string theory is the unique solution to the monodromy bootstrap with either maximal susy or vector external states. We also introduce Tachyons to the bootstrap and demonstrate for the scattering of external vectors, the bosonic and superstring span the allowed region. Allowed regions for closed string effective field theories are obtained by implementing the KLT relations.   }

\begin{document}

\maketitle

\section{Introduction}
There has been a long history of constraining effective field theories (EFT) via dispersive representations of the low energy amplitude, from pion scattering~\cite{Roy:1971tc, Ananthanarayan:1994hf, Colangelo:2001df,
Caprini:2003ta, Guerrieri:2018uew} to the sign of leading dimension eight operators~\cite{Adams:2006sv, Komargodski:2011vj, Cheung:2016yqr, deRham:2017avq}. More recently, Wilson coefficients of operators beyond dimension-eight have garnered significant interest~\cite{Guerrieri:2020bto, Tolley:2020gtv, Bellazzini:2020cot, Caron-Huot:2020cmc, Arkani-Hamed:2020blm, Sinha:2020win, Bern:2021ppb, Chowdhury:2021ynh, Guerrieri:2021ivu, Chiang:2021ziz, Caron-Huot:2021rmr, Bellazzini:2021oaj, Karateev:2022jdb, Haring:2022sdp, Caron-Huot:2022ugt, Bern:2022yes, Albert:2022oes, Guerrieri:2022sod, Albert:2023jtd}. By leveraging the dispersive representation as a tool to enforce unitarity, we can delineate the permissible parameter space for EFT coefficients. The EFT coefficients then serve as the parameters of consistent UV completions~(see Snowmass reports for review~\cite{deRham:2022hpx, Kruczenski:2022lot}). One of the main objectives is to glean universal bounds for gravitational EFTs, with the hope that information on consistent quantum gravity theories can be recovered. On the other hand, in some sense, the ``dual" question, namely how much of the theory space can be realized in string theory, remains an open question. 

To address this question in a meaningful way, we need a working definition of string theory. In the context of perturbative string theories, we can equate string theory with the four-point amplitude that admits a worldsheet CFT construction. That is, one instead asks what is the signature of a worldsheet in the S-matrix? In an earlier work by one of the authors~\cite{Huang:2020nqy}, it was suggested that for open strings, the monodromy relations for distinct ordered amplitudes can serve as a universal feature. To be concrete, for tree-level amplitudes the opens string is given by a disk correlator, where singularities can only occur when insertion points coincide. Since such singularities encode the factorization of the amplitude, its form is severely constrained. For example, if we consider an open string propagating in $R^{1,d{-}1}\otimes M$ where $M$ is some compact manifold, this implies that the four-point amplitude takes the form
\begin{equation}
A_4(1234)=J\int_{\mathbb{I}} \prod_{i=1}^4 dz_i\; |z_{i,j}|^{k_i\cdot k_j{+}a_{i,j}} f(\epsilon_i, z_i, k_i)
\end{equation}
where $(k_i,\epsilon_i)$ are the $d$-dimensional momentum and polarization tensor respectively, and $\mathbb{I}$ represents the order of the the insertions. Irrespective of the function $f$, we can immediately deduce that the amplitudes of distinct ordering must be related via monodromy relations:
\begin{eqnarray}
A_4(1324)+e^{i \pi (k_2\cdot k_3{+}a_{23}) }A_4(1234){+}e^{i \pi ({-}k_2\cdot k_4{-}a_{24})}A_4(2134)=0\,.
\end{eqnarray}
Expanding the above at low energies then implies non-trivial relations between the Wilson coefficients. Importantly, these relations involve coefficients with distinct derivative counting. For example, for the scattering of massless scalars, the low energy amplitude can be parameterized as:
\begin{equation}
        A_4(1234) =  - \left( \frac{s}{t} + \frac{t}{s} \right) + b \left( \frac{1}{s} + \frac{1}{t} \right) + \sum_{k,q \geq 0} g_{k,q} s^{k-q} t^q.
        \end{equation}
The monodromy relations (with $a_{ij}$ set to 0) then fixes the coefficients to $b=0$ and
\begin{align}\label{eq: NoneSusyEFT}
    &\begin{pmatrix}
        g_{0,0}\\
        g_{1,0} & g_{1,1}\\
        g_{2,0} & g_{2,1} & g_{2,2}\\
        g_{3,0} & g_{3,1} & g_{3,2} & g_{3,3}\\
    \end{pmatrix}
    =\begin{pmatrix}
        -1\\
        g_{1,0} & g_{1,0}\\
        \frac{\pi^2}{6} & \frac{\pi^2}{6} & \frac{\pi^2}{6}\\
        g_{3,0} & 2 g_{3,0} {-}\frac{\pi^2}{6} g_{1,0}  & 2g_{3,0} {-}\frac{\pi^2}{6} g_{1,0} & g_{3,0} \\
    \end{pmatrix}    
\end{align}
While certain coefficients are fixed, for example $g_{2,0}=\frac{\pi^2}{6}$, there are many that remains free. In~\cite{Huang:2020nqy}, it was proposed that these unfixed coefficients can be severely constrained by demanding unitarity. More precisely, unitarity demands that the Wilson coefficients reside in the EFThedron~\cite{Arkani-Hamed:2020blm}, which is a convex hull of vectors obtained from the derivative expansion of the partial wave polynomials. This is simply a geometric statement of the usual dispersive representation of the EFT couplings. Remarkably, by assuming maximal susy, the Wilson coefficients are cornered to a small region around the type-I superstring values. It was conjectured, that the Wilson coefficients of type-I superstring is a solution to an intersection geometry: that between the EFThedron and the modular-plane.

The goal of this paper is three-fold. Firstly, since the results of~\cite{Huang2019-yq}, significant progress has been made in incorporating numeric optimization to the EFT bootstrap~\cite{Caron-Huot:2020cmc, Caron-Huot:2021rmr} as well as its analytic description in terms of geometry~\cite{Arkani-Hamed:2020blm, Chiang:2021ziz}. Thus we expect to test the conjecture to higher numerical precision. In the numeric approach, one utilizes semi-definite programming (SDPB)~\cite{Simmons-Duffin:2015qma} to optimize the couplings expressed via their dispersive representations subject to constraints. In our case, these constraints are the monodromy relations.\footnote{Since we are considering ordered amplitudes on a disk, crossing symmetry is replaced by cyclic symmetry. The latter turns out the be a subset of the monodromy relations.} The result one obtains for each coupling are double-sided bounds which depend on the space-time dimensions. With maximal SUSY, we find that in general the Wilson coefficients can be fixed to within $10^{-4}$ of the superstring value. For example for $\tilde{g}_{1,0}$, $\tilde{g}_{3,0}$, $\tilde{g}_{4,1}$, which are related to the coefficients of $D^2F^4, D^6F^4, D^8F^4$, are constrained to the following:
\begin{center}
    \begin{tabular}{|c|c|c|c|c|c|}
    \hline
       D & $\tilde{g}_{k,q}$ & Two-sided bound & Superstring value & Relative error  \\
       \hline
       4 & $\tilde{g}_{1,0}$ & $1.20204774 < \tilde{g}_{1,0} < 1.20205755$ & 1.20205690 & $8.1 \times 10^{-6}$ \\
       \hline
       4 & $\tilde{g}_{3,0}$ & $1.03692704 < \tilde{g}_{3,0} < 1.03692956$ & 1.03692775 & $2.4 \times 10^{-6}$ \\
       \hline
       4 & $\tilde{g}_{4,1}$ & $0.0405367063 < \tilde{g}_{4,1} < 0.0405469176$ & 0.0405368972 & $2.5 \times 10^{-4}$ \\
       \hline
       10 & $\tilde{g}_{1,0}$ & $1.20205185 < \tilde{g}_{1,0} < 1.20205700$ & 1.20205690 & $4.3 \times 10^{-6}$ \\
       \hline
       10 & $\tilde{g}_{3,0}$ & $1.03692764 < \tilde{g}_{3,0} < 1.03692814$ & 1.03692775 & $4.8 \times 10^{-7}$ \\
       \hline
       10 & $\tilde{g}_{4,1}$ & $0.0405368583 < \tilde{g}_{4,1} < 0.0405426553$ & 0.0405368972 & $1.4 \times 10^{-4}$ \\
       \hline
    \end{tabular}
\end{center}
This lends strong support to the conjecture in~\cite{Huang2019-yq}. Notice that we did not give bounds beyond $D=10$. As it turns out, monodromy relations with maximal susy are incompatible with unitarity for $D>10$. While this is not surprising from the string theory point of view, it is remarkable that we can obtain this conclusion from the low-energy EFT with just a handful of couplings. Importantly, a numeric certificate for in-feasibility was obtained utilizing our geometric formulation of the problem, where necessary conditions for the compatibility of monodromy relations and unitarity can be stated irrespective of any truncation issues. We also consider scalar effective field theory without assuming supersymmetry. Once again, monodromy severely constrains the space of couplings, albeit no longer to isolated points, as demonstrated in the following figure 
$$\includegraphics[width = 0.7 \linewidth]{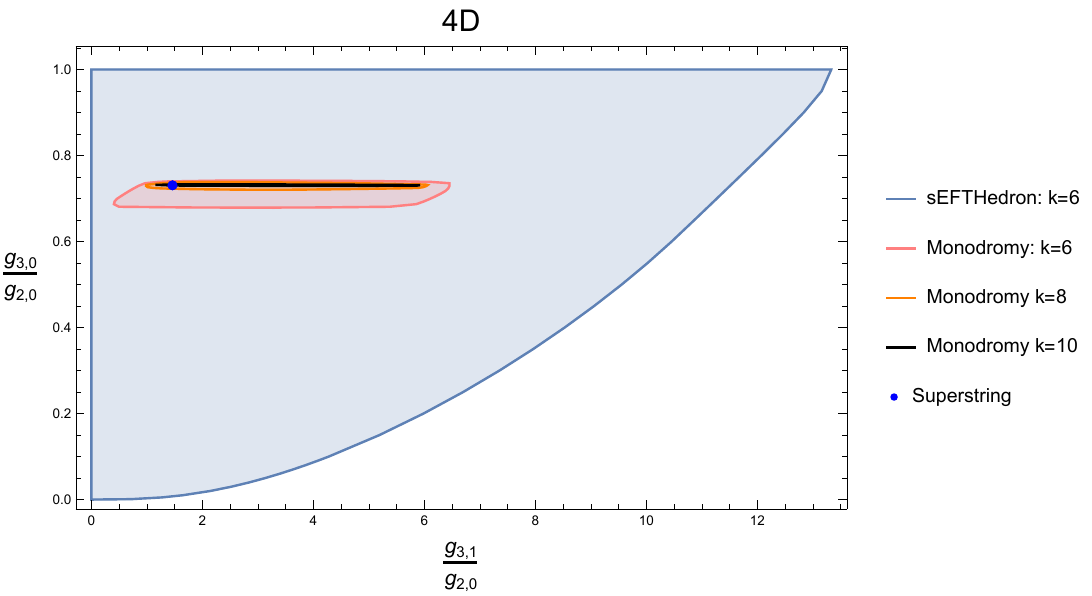}$$
We have plotted the allowed region by monodromy$+$ unitarity for $\left(g_{3,0},g_{3,1}\right)$ normalized by $g_{2,0}$, compared to unitarity alone derived in \cite{Chiang:2021ziz}. Notice that the ratios are isolated to a point in the direction $g_{3,0}/g_{2,0}$ but not $g_{3,1}/g_{2,0}$. For non-supersymmetric scalars, we can show that within current computation limits, the critical dimension is below $D=12$.

Secondly, we can apply the monodromy relations to the scattering of mass-less gauge fields. In four-dimensions, this simply implies that the dispersive representation is given in terms of Wigner d-matrices. Amazingly, we again find that the two-sided bounds closely hug the super-string value. For example for the couplings $b_{1,0}, b_{2,0}, b_{3,2}$, which corresponds to $D^2F^4, D^4F^4, D^6F^4$ respectively: 
\begin{center}
    \begin{tabular}{|c|c|c|c|c|c|}
    \hline
        $b_{k,q}$ & Two-sided bound & Superstring value & Relative error  \\
       \hline
        $b_{1,0}$ & $1.20204774 < b_{1,0} < 1.20205755$ & 1.20205690 & $8.1 \times 10^{-6}$ \\
       \hline
        $b_{2,0}$ & $1.03692704 < b_{2,0} < 1.03692956$ & 1.03692775 & $2.4 \times 10^{-6}$ \\
       \hline
        $b_{3,2}$ & $0.0405367063 < b_{3,2} < 0.0405469176$ & 0.0405368972 & $2.5 \times 10^{-4}$ \\
       \hline
    \end{tabular}
\end{center}
Deviations from the superstring only occur when Tachyons are introduced. In such case, the allowed region becomes a one-dimensional line as demonstrated in the following figure
$$\includegraphics[width = 0.7 \linewidth]{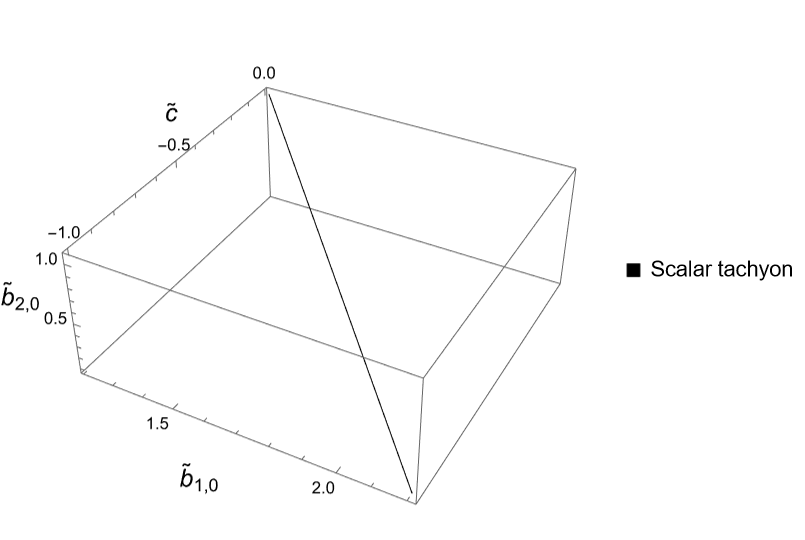}$$
where the endpoints perfectly match with the bosonic and superstring respectively. The comparison is listed in the following table 
$$\begin{tabular}{|c|c|c|}
            \hline
            Wilson coefficients & Error of bosonic string & Error of Superstring \\\hline
            $\tilde{b}_{1,0}$ &  (Max)\;$8.8\times10^{-4}$ & (Min)\;$1.6\times10^{-5}$\\\hline
            $\tilde{b}_{2,0}$ & (Min)\;$2.5\times10^{-2}$ & (Max)\;$9.2\times10^{-6}$ \\\hline
            $\tilde{b}_{3,2}$ &(Max)\;$1.8\times10^{-3}$ & (Min)\;$1.5\times10^{-3}$ \\\hline
            $\tilde{c}$ & (Min)\;$2\times10^{-3}$ &(Max)\;N/A \\\hline
        \end{tabular}
$$
where (Max/Min) indicates that it is the maximum/minimum bound that is near to the indicated string theory value. This indicates that the space of vector EFT with a scalar Tachyon in the spectrum is spanned by the bosonic string and superstring.

Finally, we consider the corresponding closed string amplitude via the KLT double copy~\cite{Kawai:1985xq}. Since the open string amplitudes satisfy monodromy relations, the resulting KLT double copy will be automatically crossing symmetric. However, unitarity is not guaranteed. Thus by considering the intersection of the KLT product and the unitarity bounds for gravitation theories, we can carve out the theory space associated with closed strings. Interestingly, all KLT products for four scalars/graviton resides in the gravitational unitarity bounds~\cite{Caron-Huot:2021rmr, Caron-Huot:2022ugt}. 

This paper is organized as follows, in the next section we begin with a brief review of the derivation of monodromy relations for disk amplitudes as well as dispersive representations for the EFT coefficients. In section~\ref{sec: susy} we first consider bootstrapping scalar EFT under the assumption of maximal susy. We discuss bounds, critical dimensions, and the spectrum of the extremal functionals. Susy is relaxed in sec~\ref{sec: None-susy}, where the same analysis is repeated. In sec~\ref{sec: Vectors} we consider the scattering of gluons instead where the dispersive representation is of different. Finally, we utilized the KLT double copy to analyze the resulting closed string EFT in sec~\ref{sec: GraviEFT}. 

$$\;$$

During the preparation of this draft, the authors became aware of upcoming work~\cite{NewPaper} which reaches a similar conclusion for the supersymmetric bootstrap. In particular, the isolation of the type-I result was established beyond $\tilde{g}_{4,1}$ and up to $\tilde{g}_{8,1}$. 

\section{Setup}
We begin with a brief review of how monodromy relations for color-ordered amplitudes arise from a worldsheet picture, as well as how the requirement of a unitary UV completion imposes constraints on the effective field theory amplitude. While the latter is applicable in a general context, the former is only established for standard type-I and bosonic string amplitudes. As we will see, under the assumption of massless poles, we can argue that the standard monodromy relations apply in a wider context.  
\subsection{Monodromy relations}
Let's consider general features of perturbative closed string amplitudes with four graviton external states. This is described by a worldsheet integral over a Riemann sphere:
\begin{equation}
M(s,t)=\int \prod_i dz_i^2\;\frac{1}{J}\; f^{\rm sphere}(z_i,\bar{z}_i, k_i,\epsilon_i)
\end{equation}
where $(k_i,\epsilon_i)$ are the polarization tensors and momenta of the external graviton, and $J$ is the Jacobian factor associated with fixing three punctures, 
\begin{equation}
J=\frac{d^2 z_a d^2 z_b d^2 z_c}{|z_{ab}|^2|z_{bc}|^2|z_{ac}|^2}
\end{equation} 
The function $f^{\rm sphere}$ is a four-point correlation function of a 2D CFT on a sphere. At this point, one can try to utilize the conformal block expansion of general four-point correlation functions and extract constraints on $M(s,t)$.

Here, we instead proceed by assuming that the background geometry takes a product form, namely $\mathbb{R}^{1,d{-}1}\otimes M$. In such case, the vertex operator factorizes as well:
\begin{equation}
\mathcal{V}=e^{i k\cdot x} V(k,x,\epsilon) U(X)
\end{equation}
where $k,x$ are the momenta and position for  Minkowski space $\mathbb{R}^{1,d{-}1}$. The closed string amplitude then takes the form 
\begin{equation}\label{eq: ClosedString}
M(s,t)=\int \prod_i dz_i^2\;\frac{1}{J} \prod_{i\neq j} |z_{ij}|^{-k_{ij}} \tilde{f}^{\rm sphere}(z_i, \bar{z}_i, k_i, \epsilon_i)
\end{equation}
The presence of the Koba-Nielsen (KN) factor $z^{-k_{ij}}_{ij}$ guarantees that the amplitude enjoys exponential suppression at high energy fixed angle~\cite{Gross:1987ar}. The function $\tilde{f}^{\rm sphere}$ is analytic in $z_i$ except for when two punctures collide. The limiting behaviour dictates the singularity structure of the amplitude. For example, writing $z_i=z_j{+}\tau e^i\theta$ near the insertion, the contribution of the worldsheet integral near the insertion point yields 
\begin{equation}
\int d\tau \tau^{-k_{ij}{+}\alpha_{ij}} \sim \frac{1}{-k_{ij}{+}\alpha_{ij}{+}1}\,.
\end{equation}
where $\alpha_{ij}$ represent contributions from $\tilde{f}$. Thus we immediately arrive at the conclusion that if the amplitude is to have a massless pole in a given channel, say $k_i\cdot k_j$, then $\alpha_{ij}$ is at most an integer. Thus by requiring the presence of massless poles in all channels, as one would expect for graviton amplitude, $\tilde{f}^{\rm sphere}$ must have trivial monodromies.\footnote{None-rational exponents can produce usual propagator singularities if one sums over different contributions. For example 
$$\sum_\pm \frac{1}{m\pm \sqrt{s}}=\frac{2m^2}{m^2-s}$$
However, it is not clear how worldsheet Green's functions can produce such exponents. See ~\cite{Cheung:2023uwn} for related discussions.}   

At this point, it is still not obvious how to extract constraint on $M(s,t)$ from the representation in eq.(\ref{eq: ClosedString}). However, since $\tilde{f}^{\rm sphere}$ has trivial monodromy, we can straightforwardly apply the KLT contour deformation~\cite{Kawai:1985xq} (see ~\cite{Sondergaard:2011iv} for review) to rewrite:
\begin{equation}\label{eq: KLT}
M(s,t)=\sin \pi s A(s,t)A(s,u)
\end{equation}
where $A(s,t)$ is the open string amplitude of four massless states. Fixing the previous punctures $z_i$to $0,z,1,\infty$ respectively, we arrive at the amplitude 
\begin{equation}
A(s,t)=\int_{\mathbf{I}} dz|z|^{k_1\cdot k_2} |z-1|^{k_2\cdot k_3} f^{\rm disk}(z,\epsilon_i, k_i)
\end{equation}
The integration domain $\mathbf{I}$ now depends on the ordering of the amplitude. For example for $A(s,t)$ we have $0<z<1$. Note that the different integration domain does not affect the form of $f^{\rm disk}$, only the KN factor due to the absolute value. More precisely we have: 
\begin{eqnarray}
A(1234)&\sim&  \int_{0}^1 dz \; z^{k_{12}} (1-z)^{k_{23}}f^{\rm disk}\nonumber\\
A(1324)&\sim& \int_{1}^\infty dz\; z^{k_{12}} (z-1)^{k_{23}}f^{\rm disk}\nonumber\\
A(2134)&\sim& \int_{{-}\infty}^0 dz\; ({-}z)^{k_{12}} (1-z)^{k_{23}}f^{\rm disk}\nonumber\\
\end{eqnarray}
The different ordered amplitudes can be related by monodromy relations~\cite{Stieberger:2009hq, Bjerrum-Bohr:2009ulz}, which we now review. 

Begin with the amplitude $A(u,t)$, where the integration region is between $1$ and $+\infty$ 
$$\includegraphics[scale=0.6]{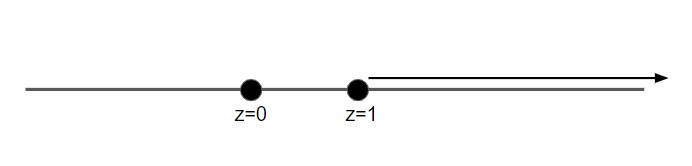}$$
We can deform the contour such that it becomes 
$$\includegraphics[scale=0.6]{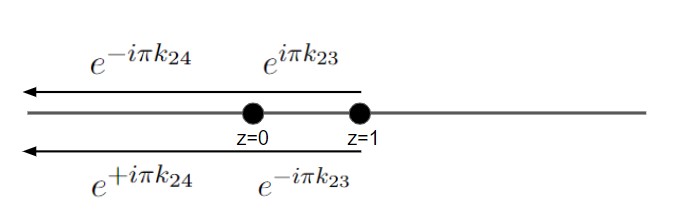}\,.$$
Note that since all the singularities are at the boundary of the disk, the deformation does not pick up any poles. 
Now due to the non-rational exponents, a deformation from the above and below will attain different monodromy,
\begin{eqnarray}
z^{k_{12}} (z-1)^{k_{23}} &\rightarrow& \begin{matrix}
    z^{k_{12}} (1-z)^{k_{23}}e^{i \pi k_{23} }\\
    z^{k_{12}} (1-z)^{k_{23}}e^{-i \pi k_{23} }
\end{matrix}\nonumber
\rightarrow \begin{matrix}
    (-z)^{k_{12}} (1-z)^{k_{23}}e^{-i \pi k_{24} }\\
    (-z)^{k_{12}} (1-z)^{k_{23}}e^{+i \pi k_{24} }
\end{matrix}\,.
\end{eqnarray}
Thus we immediately have the identity 
\begin{eqnarray}\label{eq: MonodromyRel}
A(u,t)&=&{-}\mathbf{Re}\left[e^{i \pi k_{23} }A(s,t){+}e^{{-}i \pi k_{24}}A(s,u)\right] \nonumber\\
0&=&\mathbf{Im}\left[e^{i \pi k_{23} }A(s,t){+}e^{-i \pi k_{24}}A(s,u)\right]\,. 
\end{eqnarray}
 The above monodromy relations become extremely powerful as one considers the low energy limit, whereby expanding in the Mandelstam variables, eq.(\ref{eq: MonodromyRel}) leads to linear relations amongst EFT couplings. Importantly, these relations combine couplings of \textit{different} mass dimensions! Thus such monodromy relations are the IR avatar of an underlying worldsheet. 

\subsection{Unitarity}
Unitarity is reflected in the positivity of the imaginary part of partial waves that appear in the dispersive representation of the scattering amplitude. As we are considering ordered amplitude, for fixed $t$, the imaginary part arises from threshold production in only one channel. For $A(s,t)$, we will only have $s$-channel non-analyticity, leading to the following dispersive representation 
\begin{equation}
A(s,t)=-\int_{M^2}^\infty ds'\;\frac{\rho_\ell(s')G^D_{\ell}(1+2t/s')}{s-s'}
\end{equation}
where $M^2$ is the scale of the first resonance and the equality is understood in terms of matching the Taylor coefficients of $s^{k-q}t^q$ on both sides, with $(k{-}q)\geq2$. Here $\rho_\ell={\rm Im}[a_\ell]\geq0$ where $a_\ell$ are the partial wave coefficients of the amplitude. The fact that the equality holds for $(k{-}q)\geq2$ is due to the fact that in general, we expect the amplitude to be bounded by $A(s,t)<s^2$, and thus it satisfies twice subtraction. 

In a nutshell, unitarity tells us that the EFT couplings $g_{k,q}$ must reside in a convex hull, where elements in the hull are labeled by the mass and the spin. For convenience from now on we will term the space spanned by this hull the ``EFThedron", following~\cite{Arkani-Hamed:2022gsa}. At the same time, these couplings must reside on a hyper-plane defined by the linear relations implied by the monodromy relations. Thus the allowed region for each coupling is given by the intersection of the monodromy plane and the EFThedron.   

There are two challenges in characterizing this region. First, is that invariantly we will be considering a finite number of couplings, and this restricts our access to the number of monodromy relations. Thus in the intersecting geometry picture, we are considering the limit where the couplings beyond the truncation are projected out. This leads to weaker bounds since a non-intersecting configuration in higher dimensions can be projected to intersecting ones at lower dimensions.  Second, there is in principle an infinite number of elements in the convex hull, and thus the boundary of the EFThedron is difficult to compute. In the following, we will proceed to analyze the problem using two complementary approaches:  

\begin{itemize}
    \item \textbf{Numerical SDPB approach}: Starting with the low energy amplitude, say $A(s,t)$, we give the dispersive representation for each coupling. In general, we have
   \begin{equation}
   g_{k,q}=\frac{1}{q!}\frac{\partial^q}{\partial t^q}\int \frac{ds'}{s'^{k-q+1}} \sum_{\ell} \rho_{\ell }(s')G^D_{\ell}(1+2t/s')\bigg|_{t=0}
   \end{equation}
   Then, substituting these dispersive representations into the linear relations in eq.(\ref{eq: MonodromyRel}) gives the null constraints on the UV sum rules. After a suitable rescaling of the mass, the UV sum rules is over a functions of spin and mass. This allows us to utilize SDPB~\cite{Simmons-Duffin:2015qma} to find functionals comprised of linear combinations of null constraints such that it is positive definite over arbitrary spins and mass and give optimal bounds on the couplings. Note that in general, one needs to truncate the sum of partial waves to some maximal spin $\ell=\ell_{max}$, i.e. spin truncation. The presence of spin-truncation makes the SDPB problem an approximation of the original problem.  Thus the result corresponds to a \textit{valid bound}, for a fixed number of null constraints, \textit{only if convergence with respect to $\ell_{max}$ can be established}.

    \item \textbf{Generalized Hankel matrix approach}: the EFThedron can be identified with a simple mathematical object~\cite{Arkani-Hamed:2020blm, Chiang:2021ziz}: the convex hull of a double moment $\mathbb{R}^+\otimes \mathbb{Z}^+$. Finding the boundary of such space constitutes the well-known ``bi-variate moment problem", which seeks to find sufficient conditions for a point to be inside the hull. For the ``singe moment problem", the solution is known. Here one asks for sufficient conditions on $\vec{y}$ such that 
    \begin{equation} 
    \vec{y}=\int_I \rho(x) \vec{x},\quad  \begin{pmatrix}
            1 \\
            x \\
            x^2 \\
            \vdots
        \end{pmatrix}
    \end{equation}
  for some $\rho(x)>0$ and $I$ is the domain of $x$. The solution is that the Hankel matrix 
  \begin{equation}
  \mathcal{H}[\vec{y}]=\begin{pmatrix}
            y_0 &y_1& y_2 &\cdots \\
            y_1 & y_2 & y_3 &\cdots \\
            \vdots & \vdots & \vdots & \; 
        \end{pmatrix}=\int_I\, \rho(x) \vec{x}\vec{x}^T\geq 0\,,
  \end{equation}
  is positive semi-definite. The inequalities associated with the above condition then constitute the boundary of the hull. For the bi-variate moment problem a general solution is not known. However, one can easily write down a set of \textit{necessary conditions} by invoking the positivity of the ``generalized Hankel matrix" which we will introduce in eq.(\ref{eq: vanilla_Hankel}) and eq.(\ref{eq: shift}). As the couplings must reside on the monodromy plane, the Hankel matrices are parameterized by the coordinates on the plane. Thus one simply searches for the maximum/minimum of a given coordinate that yields a positive semi-definite Hankel matrix, which is a standard semi-definite programming SDP problem. Such an approach has been utilized in the context of EFT~\cite{Chiang:2021ziz, Chiang:2022ltp, Chiang:2022jep} as well as modular bootstrap~\cite{Chiang:2023qgo}. 
 
 While the positivity of generalized Hankel matrices are not sufficient conditions, the fact that they are necessary implies we can use them to rigorously determine \textit{when} the geometry does not intersect. This will become useful in obtaining the critical dimension as we will see.    
\end{itemize}

\section{Maximal SUSY}\label{sec: susy}
Let us first consider the effects of monodromy relations in eq.(\ref{eq: MonodromyRel}) for an EFT with maximal susy. In such case, the low energy amplitude takes the form

 \begin{equation}
     \quad \mathcal{A}(s,t){=}\delta^8(Q)f(s,t)=\delta^8(Q)\left[\frac{-1}{st} + b \left(\frac{1}{s} + \frac{1}{t} \right) + c \left(\frac{t}{s} + \frac{s}{t} \right) + \sum_{k,q \geq 0} \tilde{g}_{k,q} s^{k-q} t^q\right].
    \end{equation}
The monodromy relations will immediately set $b=0$. In fact, since the coefficients $b,c$ represent the contribution from operators $F^2\phi$ and $F^3$ respectively, we will set both of them to zero as they are forbidden by SUSY. With this setup, the monodromy relations up to $k=4$ take the form:
\begin{align}\label{eq: MonodromySUSY}
    &\begin{pmatrix}
        \tilde{g}_{0,0}\\
        \tilde{g}_{1,0} & \tilde{g}_{1,1}\\
        \tilde{g}_{2,0} & \tilde{g}_{2,1} & \tilde{g}_{2,2}\\
        \tilde{g}_{3,0} & \tilde{g}_{3,1} & \tilde{g}_{3,2} & \tilde{g}_{3,3}\\
        \tilde{g}_{4,0} & \tilde{g}_{4,1} & \tilde{g}_{4,2} & \tilde{g}_{4,3} & \tilde{g}_{4,4}\\
    \end{pmatrix} = \begin{pmatrix}
        \frac{\pi^2}{6}\\
        \tilde{g}_{1,0} & \tilde{g}_{1,0}\\
        \frac{\pi^4}{90} & \frac{\pi^4}{360} & \frac{\pi^4}{90}\\
        \tilde{g}_{3,0} & -\frac{\pi^2}{6} \tilde{g}_{1,0} + 2 \tilde{g}_{3,0} & -\frac{\pi^2}{6} \tilde{g}_{1,0} + 2 \tilde{g}_{3,0} & \tilde{g}_{3,0}\\
        \frac{\pi^6}{965} & \tilde{g}_{4,1} & -\frac{\pi^6}{15120} + 2 \tilde{g}_{4,1} & \tilde{g}_{4,1} & \frac{\pi^6}{965}
    \end{pmatrix}
\end{align}
Note that the cyclic symmetry of the disk amplitude is automatically implied by the monodromy relations. 

Next, we express the couplings through its dispersive representation. With maximal SUSY, we can choose the scattered states such that $\delta^8(Q)\sim s^2$.\footnote{Considering other arrangements for the external states does not lead to new constraints. As discussed in~\cite{Arkani-Hamed:2022gsa}, once the imaginary part of $f(s,t)$ has a positive expansion in terms of scalar Gegenbauers, this is a sufficient condition for unitarity.} In such case we expect that
\begin{equation}
f(s,t)|_{s\rightarrow \infty}\sim s^0\,.
\end{equation}
This implies that $f(s,t)$ admits a zero subtraction dispersive representation. That is, 
\begin{equation}
f(s,t)=\int_{M^2}^\infty ds' \;\frac{\sum_\ell\rho_\ell(s')G^{D}_\ell(1+2t/s')}{s'{-}s}
\end{equation}
where the equality is again in terms of Taylor expansion in $s,t$, and $G^{D}_\ell$ is the $D$-dimensional Gegenbauer polynomial. The spectral function $\rho_\ell$ is positive due to unitarity. Thus we have 
\begin{equation}\label{gkq_disp}
g_{k,q}=\int_{M^2}^\infty \frac{ds'}{(s')^{k{+}1}}\;\sum_\ell\rho_\ell(s') v^{D}_{\ell,q}
\end{equation}
where $v^{D}_{\ell,q}$ is the coefficient of $t^q$ in the Taylor expansion of  $G^{D}_\ell(1+2t)$. Using the dispersive representation for the LHS of eq.(\ref{eq: MonodromySUSY}) and equating to the right then constitutes the ``null" constraints. We then find the maximal and minimal values of each unfixed Wilson coefficient subject to a number of null constraints up to $k=k_{\rm max}$. 

\subsection{EFT space}
As shown in eq.(\ref{eq: MonodromySUSY}), there are three unfixed coefficients up to $k=4$, $\tilde{g}_{1,0},\, \tilde{g}_{3,0},\, \tilde{g}_{4,1}$ which corresponds to operators $D^2F^4,\,D^6F^4,\,D^8F^4$ operators respectively. We first apply SDPB with fixed $(D,\,k_{\rm max},\,\ell_{\rm max})$, where $\ell_{\rm max}$ is the spin truncation. We extract upper (lower) limits for each coupling, which monotonically decreases (increases)  in $(k_{\rm max},\,\ell_{\rm max})$. 

\paragraph{Numerical SDPB setup} Bounding the EFT Wilson coefficients using SDPB has been done in several works, including~\cite{Caron-Huot:2020cmc,Chiang:2021ziz}. Here we give a lightning review of this approach. Given the dispersion relation for the Wilson coefficients, i.e. eq. (\ref{gkq_disp}), we want to extremize some particular $g_{k,q}$ under the constraint of monodromy relations (\ref{eq: MonodromySUSY}). That is, we have the following optimization problem:
\begin{align}\label{sdpb_example}
    &\text{Minimize} \quad  g_{k,q} = \sum_{\ell\in\mathcal{S}}\int_{M^2}^{\infty} ds'\rho_{\ell}(s')G_{k,q,\ell}(s')\nonumber \\
    &\text{subject to} \quad  n^{(\alpha)} =\sum_{\ell\in \mathcal{S}}\int_{M^2}^{\infty} ds'\rho_{\ell}(s')N^{(\alpha)}_{\ell}(s') = c^{(\alpha)},
\end{align}
where $\alpha = 1, ..., P$ labels the monodromy constraints and $c^{\alpha}$ is some numerical constant. The functions $G_{k,q,\ell}(s)$ and $N_{\ell}(s)$ are both polynomials in $s$\footnote{This can always be done by absorbing powers of $s$ into the spectral function $\rho_{\ell}(s)$ in eq.(\ref{gkq_disp}). Positivity of the spectral function is preserved since $s > 0.$}. This can be transformed into the following program:
\begin{align}
    & \text{Maximize} \quad -\vec{c} \cdot \vec{z} \nonumber\\
    &\text{subject to} \quad \vec{z}\cdot\vec{N}_{\ell}(s')+G_{k,q,\ell}(s')\geq 0\quad \forall s'\geq M^2 \text{ and } \ell \in \mathcal{S},
\end{align}
where $\mathcal{S}$ consists of all the allowed spins of the UV states and 
$
    \vec{N}_{\ell} = \left(
        N_{l}^{(1)}, ...,
        N_{l}^{(P)}
    \right)
$
is the collection of monodromy constraints. In practice, one must impose a spin truncation $\ell_{max}$ to make the SDPB problem well-defined.

One sees that every feasible solution $\vec{z}$ of the problem provides a lower bound of the original problem eq.(\ref{sdpb_example}):
\begin{equation}
    g_{k,q} = \sum_{\ell\in \mathcal{S}}\int_{M^2}^{\infty} ds'\rho_{\ell}(s')G_{k,q,\ell}(s')\nonumber \geq - \vec{z} \cdot \sum_{\ell \in\mathcal{S}}\int_{M^2}^{\infty} ds'\rho_{\ell}(s') \vec{N}_{\ell}(s') = -\vec{c} \cdot \vec{z}.
\end{equation}
Maximizing $-\vec{c} \cdot \vec{z}$ therefore gives the strongest bound on $g_{k,q}$ as desired.

\paragraph{The EFT space} For $D^2F^4$ the resulting bounds are plotted below. We show the maximum and minimum values under up to $k=8$ and $k=10$ monodromy constraints. For $k=8$, there are 37 monodromy constraints, and for $k=10$, there are 54 constraints. We impose unitarity at various dimensions $D$ and show the result at the $\ell_{\rm max}$ where convergence has been established. 
$$
    \includegraphics[width = 0.9 \linewidth]{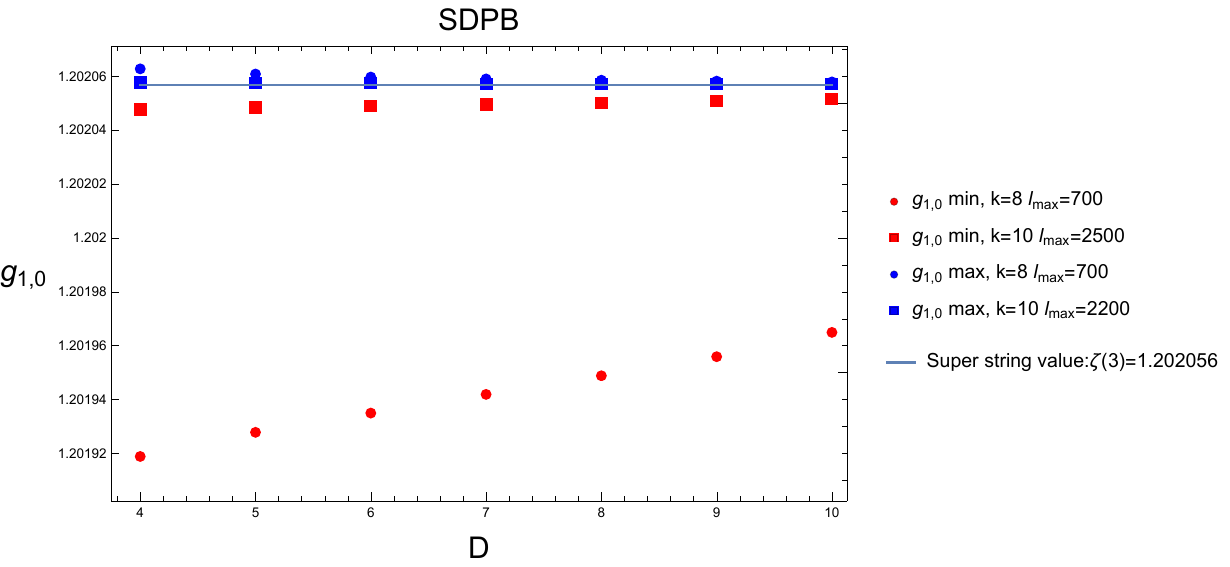}$$
    We see that while the maximum value is close to the Type-I value at $\zeta(3)=1.202056$, the minimum for $\tilde{g}_{1,0}$ is slightly off. However, the gap rapidly closes as we increase the number of null constraints from $k=8$ to $k=10$. As a result, for $D=10$ we find ${\rm min}\,\tilde{g}_{1,0}=1.20205185$ and ${\rm max}\,\tilde{g}_{1,0}=1.20205700$. This gives a region whose relative error to the type-I value, defined as $({\rm max}\,\tilde{g}_{1,0}-{\rm min}\,\tilde{g}_{1,0})/\tilde{g}^{\rm type{-}I}_{1,0}$ as $4.3 \times 10^{-6}$! 

\begin{figure}[h]
    \centering
    \includegraphics[width = 0.9 \linewidth]{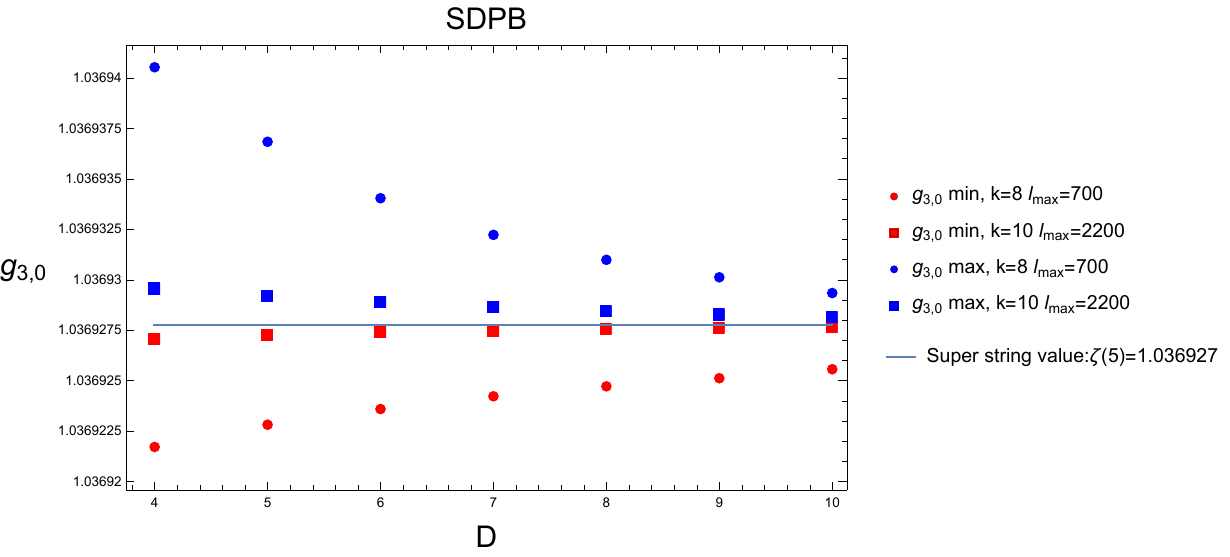}
    \caption{Upper and lower bound of $g_{3,0}$ at various space-time dimension. The results of imposing $k=8$ and $k=10$ monodromy constraints are shown in the graph. The bounds are stabilized against spin truncation.}
    \label{fig: susy_g30_dimensionplot}
\end{figure}

\begin{figure}[h]
    \centering
    \includegraphics[width = 0.9 \linewidth]{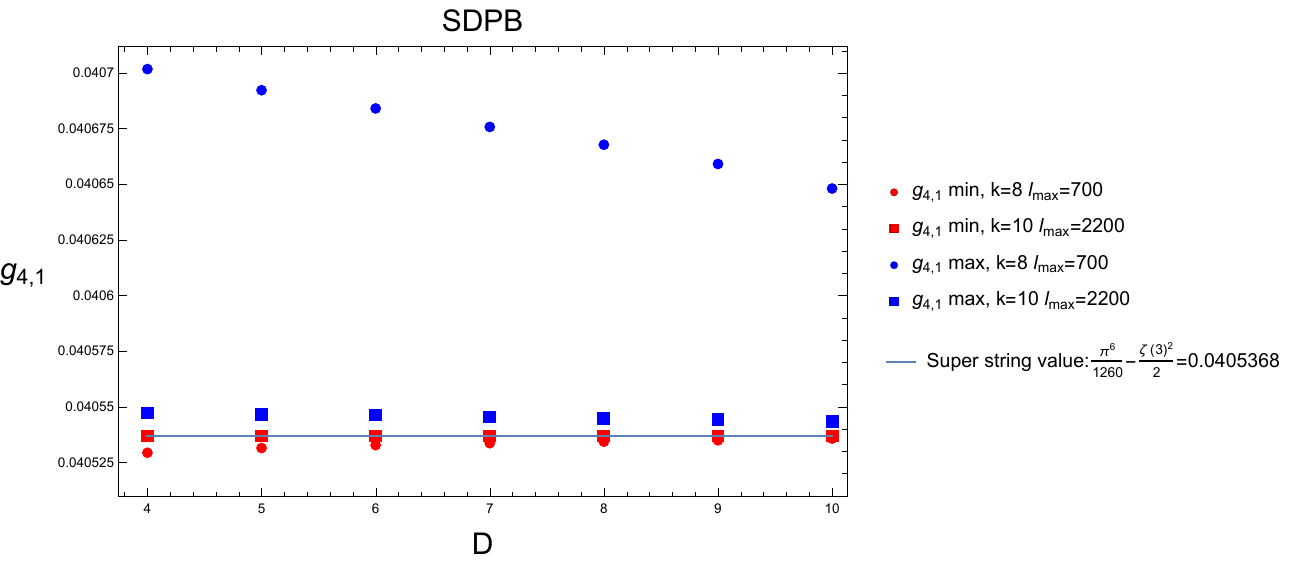}
    \caption{Upper and lower bound of $g_{4,1}$ at various space-time dimension. The results of imposing $k=8$ and $k=10$ monodromy constraints are shown in the graph. The bounds are stabilized against spin truncation.}
    \label{fig: susy_g41_dimensionplot}
\end{figure}

    Similar results for $\tilde{g}_{3,0}$ and $\tilde{g}_{4,1}$ and shown in fig.\ref{fig: susy_g30_dimensionplot} and fig.\ref{fig: susy_g41_dimensionplot} respectively. We summarize the result at $\ell_{\rm max}=2200$ in the following table
$$
    \begin{tabular}{|c|c|c|c|c|c|}
    \hline
       D & $\tilde{g}_{k,q}$ & Two-sided bound & Superstring value & Relative error  \\
       \hline
       4 & $\tilde{g}_{1,0}$ & $1.20204774 < \tilde{g}_{1,0} < 1.20205755$ & 1.20205690 & $8.1 \times 10^{-6}$ \\
       \hline
       4 & $\tilde{g}_{3,0}$ & $1.03692704 < \tilde{g}_{3,0} < 1.03692956$ & 1.03692775 & $2.4 \times 10^{-6}$ \\
       \hline
       4 & $\tilde{g}_{4,1}$ & $0.0405367063 < \tilde{g}_{4,1} < 0.0405469176$ & 0.0405368972 & $2.5 \times 10^{-4}$ \\
       \hline
       10 & $\tilde{g}_{1,0}$ & $1.20205185 < \tilde{g}_{1,0} < 1.20205700$ & 1.20205690 & $4.3 \times 10^{-6}$ \\
       \hline
       10 & $\tilde{g}_{3,0}$ & $1.03692764 < \tilde{g}_{3,0} < 1.03692814$ & 1.03692775 & $4.8 \times 10^{-7}$ \\
       \hline
       10 & $\tilde{g}_{4,1}$ & $0.0405368583 < \tilde{g}_{4,1} < 0.0405426553$ & 0.0405368972 & $1.4 \times 10^{-4}$ \\
       \hline
    \end{tabular}
$$
The above result leads us to conclude that the monodromy relations plus unitarity uniquely selects the Type-I superstring as the only solution for maximal supersymmetry. Notice that we've only implemented unitarity up to $D=10$. For $D>10$, SDPB no longer gives bounds. While it is likely that no solution exists, we cannot establish this via SDPB since it only gives valid statements when spin-convergence is established, and if there are no bounds there's no convergence to speak of. To this end, we turn to our Hankel approach. 

\subsection{Critical dimensions and extremal spectrum}
We've seen that Wilson coefficients for the maximal supersymmetric EFT are confined to a tiny sliver surrounding the type-I superstring value. This prompts us to consider the critical dimension and analyze the spectrum associated with the extremal functionals. 

\noindent\textbf{Critical dimensions }

To find the critical dimension amounts to finding that a given set of monodromy relations are incompatible with unitarity in some given dimensions. Note that if a theory is unitary in $D$ dimensions, then it is guaranteed to be unitary in any $D'<D$.  We consider the geometric approach, where the Wilson coefficients are linearly transformed to live in the convex hull of moment curves, i.e. the unitary polytope.  The spacetime dimension comes into play as a parameter within the GL transform. Consequently, it becomes possible to exclude spacetime dimensions exceeding a critical threshold, denoted as $D_{\text{crit}}$, if one can certify the absence of an intersection between the unitary polytope and monodromy plane at $D_{\text{crit}} + 1$. 

\paragraph{The unitary polytope} It can be shown that through a GL transformation on the Wilson coefficients, the dispersion relations linearly combine into a convex hull of product moment curves. For a more detailed discussion of moment problems see~\cite{Chiang:2023qgo}
\begin{itemize}
    \item Moment geometry: the dispersion relations for the Wilson coefficients $\tilde{g}_{k,q}$ can be linear transformed into the form
    \begin{equation}\label{eq: GL}
        a_{k,q} = \sum_{\ell}\int d m^2 \rho_{\ell}(m^2) \frac{J^{2q}}{m^{2(k{+}1)}}, \quad J^2 \equiv \ell(\ell+D-3)\,.
    \end{equation}
    The linear map, which, in the first few derivative orders in $t$, is given by
    \begin{equation}
        \begin{pmatrix}
            a_{k,0} \\
            a_{k,1} \\
            a_{k,2} \\
            a_{k,3} 
        \end{pmatrix}
        =
        \begin{pmatrix}
            1 & 0 & 0 & 0 \\
            0 & \frac{2}{D-2} & 0 & 0 \\
            0 & \frac{-2}{D} & \frac{2}{D(D-2)} & 0 \\
            0 & \frac{8(D-1)}{3D(D+2)} & \frac{4(4-3D)}{3 D (D^2-4)} & \frac{4}{3 D (D^2 - 4)}
        \end{pmatrix}
        \begin{pmatrix}
            \tilde{g}_{k,0} \\
            \tilde{g}_{k,1} \\
            \tilde{g}_{k,2} \\
            \tilde{g}_{k,3} 
        \end{pmatrix}.
    \end{equation}
    Importantly the RHS of eq.(\ref{eq: GL}) constitute a product moment curve 
    \begin{equation}
    \begin{pmatrix}
            a_{0,0} &a_{0,1} & a_{0,2}&\cdots \\
            a_{1,0} &a_{1,1} & a_{1,2}&\cdots \\
            a_{2,0} &a_{2,1} & a_{2,2}&\cdots \\
            \vdots  & \vdots & \vdots & \;
        \end{pmatrix}=\sum_{\ell}\int d m^2 \rho_{\ell}(m^2) \begin{pmatrix}
            \frac{1}{m^2} \\ \frac{1}{m^4} \\ \frac{1}{m^6} \\ \vdots 
        \end{pmatrix}
        \begin{pmatrix}
            1 \\ J^2 \\ J^4 \\ \vdots 
        \end{pmatrix}^T\,.
    \end{equation}
    where the two moments take value in $\mathbb{R}^+\otimes \mathbb{R}^+$. Note that at this point, we let the spin be any real positive number. Discreteness will be imposed geometrically shortly. 
    \item Unitarity: positivity of the spectral function implies that the following generalized Hankel matrix is positive semi-definite:
    \begin{eqnarray}\label{eq: vanilla_Hankel}
        \mathcal{H}_{0,0} = \sum_\ell \int d m^2 \rho_{\ell}(m^2) 
        \begin{pmatrix}
            1 \\ x \\ x y_\ell \\ x^2 \\ x^2 y_\ell \\ x^2 y_\ell^2 \\ \vdots
        \end{pmatrix}
        \begin{pmatrix}
            1 \\ x \\ x y_\ell \\ x^2 \\ x^2 y_\ell \\ x^2 y_\ell^2 \\ \vdots
        \end{pmatrix}^T
        =
        \begin{pmatrix}
            a_{0,0} & a_{1,0} & a_{1,1} & \cdots \\ 
            a_{1,0} & a_{2,0} & a_{2,1} & \cdots \\ 
            a_{1,1} & a_{2,1} & a_{2,2} & \cdots \\ 
            \vdots & \vdots & \vdots & \ddots
        \end{pmatrix}
        \geq 0, 
    \end{eqnarray}
    where $x \equiv 1 / m^2$ and $y_\ell = J^2$. Note that we denote $\mathcal{H}_{k,q}$ as the generalized Hankel matrix with the first element being $a_{k,q}$.
    \item No tachyon: negative mass states are forbidden, i.e. $x_a > 0$, which translates into the positivity of the shifted Hankel matrix:
    \begin{equation}\label{eq: shift}
        \mathcal{H}_{1,0} = \sum_\ell \int d m^2 \rho_{\ell}(m^2) x 
        \begin{pmatrix}
            1 \\ x \\ x y_\ell \\ x^2 \\ x^2 y_\ell \\ x^2 y_\ell^2 \\ \vdots
        \end{pmatrix}
        \begin{pmatrix}
            1 \\ x \\ x y_\ell \\ x^2 \\ x^2 y_\ell \\ x^2 y_\ell^2 \\ \vdots
        \end{pmatrix}^T
         = 
        \begin{pmatrix}
            a_{1,0} & a_{2,0} & a_{2,1} & \cdots \\ 
            a_{2,0} & a_{3,0} & a_{3,1} & \cdots \\ 
            a_{2,1} & a_{3,1} & a_{3,2} & \cdots \\ 
            \vdots & \vdots & \vdots & \ddots
        \end{pmatrix} \geq 0\,.
    \end{equation}
    \item Gap state: the existence of the lightest state in the UV spectrum is just the statement $x_a \leq 1/ M_{\text{gap}}^2$. We may conventionally normalize the mass gap to unity, which implies
    \begin{equation} \label{gap}
        \mathcal{H}_{0, 0} - \mathcal{H}_{1, 0} \geq 0.
    \end{equation}
    Notice that eq.(\ref{eq: shift}) and eq.(\ref{gap}) automatically imply eq.(\ref{eq: vanilla_Hankel}).
    \item Integer spin: the spins take non-negative integer values, i.e., 
    \begin{align}
        y_\ell^2 &\geq 0, \nonumber\\ 
        \left(y_\ell - \ell(\ell+D{-}3)\right)\left(y_\ell - (\ell{+}1)(\ell{+}D{-}2)\right) &\geq 0, \quad \ell= 0, 1, 2, \cdots
    \end{align}
    resulting in
    \begin{align}
        \mathcal{H}_{2,1} & \geq 0, \label{positive_spin} \\
        -\mathcal{H}_{2,2} {+} \left(\ell(\ell+D{-}3) + (\ell{+}1)(\ell{+}D{-}2)\right)\mathcal{H}_{2,1} \quad\quad &\nonumber\\
        - \ell(\ell{+}1)(\ell{+}D{-}3)(\ell{+}D{-}2) \mathcal{H}_{2,0} & \geq 0, \quad \ell = 0, 1, 2,\cdots \,,\label{integer_spin}
    \end{align}
    Eq.(\ref{positive_spin}) imposes positive spin, and eq.(\ref{integer_spin}) imposes integer spin. 
\end{itemize}

\paragraph{Numerical SDP setup} These constraints, along with any linear function in the Wilson coefficients subject to optimization, form a semi-definite program (SDP) that can be solved using established SDP solvers, such as SDPJ \cite{fishbonechiang}. In its most general form, an SDP can be expressed as: 
\begin{align}
  \label{SDP_standard_form}
  \text{Min } & \quad \mathbf{c}^T \mathbf{x}  \\
  \label{SDP_constraints_1}
  \text{subject to } & \quad \sum_{i = 1}^{m} x_i A_i^{(l)} - C^{(l)} \succeq 0, \quad l = 1, ..., L, \\
  \label{SDP_constraints_2}
  & \quad B^T \mathbf{x} = \mathbf{b}\,, 
\end{align}
where
\begin{equation}
  \mathbf{x} \in \mathbb{R}^m, \quad A_i^{(l)} \in \mathbb{S}^{k^{(l)}}, \quad B \in \mathbf{R}^{m \times n}, \quad b \in \mathbf{R}^{n}\,.
\end{equation}
For our purpose, the vector $\mathbf{x}$ consists of moment variables $a_{k,q}$ up to certain truncated derivative order $k_{\text{max}}$, and the matrices $A_{i}^{(l)}$ are chosen so that (\ref{SDP_constraints_1}) imposes Hankel matrix positivity, where $L$ is the total number of positive matrices. Eq.(\ref{SDP_constraints_2}) represents the linear relations that the variables have to satisfy to live on the monodromy plane. In our case, these equality constraints can be easily solved to reduce the number of variables, thereby enhancing computational efficiency, i.e. we can choose a parametrization with monodromy relations manifest.

To demonstrate the absence of an intersection between the unitary polytope (\ref{SDP_constraints_1}) and the monodromy plane (\ref{SDP_constraints_2}), one minimizes an auxiliary variable $t$ that measures the violation of positivity constraints as follows:
\begin{align}
  \text{Min} & \quad t \\
  \text{s.t.} & \quad \sum_{i = 1}^{m} x_i A_i^{(l)} - C^{(l)} + t I \succeq 0, \quad l = 1, ..., L, \\
  & \quad B^T \mathbf{x} = \mathbf{b},
\end{align}
and the positivity of the optimal value $t^{*}$ will serve as a certificate of infeasibility.

\paragraph{Numerical result} The constraints (\ref{eq: vanilla_Hankel}), (\ref{eq: shift}), and (\ref{positive_spin}) were considered. With the parameters specified in Table \ref{parameters}, the results can be found in Table \ref{crit_dim_results}. Remarkably, we found a certificate of infeasibility at $k_{\text{max}} = 10$ in $D = 11$, \emph{establishing $D_{\text{crit}} = 10$ for the critical dimension!} Notably, this result was obtained by imposing only the positive spin constraint (\ref{positive_spin}) but not the integer spin. Therefore, this approach is rigorous up to the primal/dual residue shown in Table \ref{parameters} and is free from any spin-truncation issues. In a later section, we shall see that similar numerical analysis on the feasibility problem, again, sets bounds on the spacetime dimension even \emph{without} assuming SUSY.

\begin{table}[h]
  \center
  \begin{tabular}{|l|l|}
    \hline
    \texttt{beta}             &   $0.01$                  \\
    \texttt{Omega\_p}         &   $10^{10} \sim 10^{30}$\\
    \texttt{Omega\_d}         &   $10^{10} \sim 10^{30}$\\
    \texttt{epsilon\_gap}     &   $10^{-10}$\\
    \texttt{epsilon\_primal}  &   $10^{-200}$              \\
    \texttt{epsilon\_dual}    &   $10^{-200}$ \\
    \texttt{prec}             &   $300$ \\
    \hline
  \end{tabular}
  \caption{Choice of parameter for \texttt{SDPJ}. See \cite{fishbonechiang} for the definition of the parameters.}
  \label{parameters}
\end{table}

\begin{table}[h]
\center
\begin{tabular}{|l|l|l|l|l|}
 \hline
 & $D = 10$ & $D = 11$ & $D = 12$ & $D = 13$
 \\
 \hline
 $k_{\text{max}} = 6$ & $t^* < -1.35 \times 10^{-4}$ & $t^* < -8.45 \times 10^{-5}$ & $t^* < -7.05 \times 10^{-6}$ & $t^* < -5.51 \times 10^{-5}$
 \\
 \hline
 $k_{\text{max}} = 8$ & $t^* < -1.32 \times 10^{-7}$ & $t^* < -2.64 \times 10^{-7}$ & \color{red}{$t^{*} > 3.70 \times 10^{-5}$} & \color{red}{$t^{*} > 3.19 \times 10^{-4}$}\\
 \hline
 $k_{\text{max}} = 10$ & $t^* < -5.53 \times 10^{-9}$ & \color{red}{$t^{*} > 8.62 \times 10^{-5}$} & \color{red}{$t^* > 3.43 \times 10^{-4}$} & \color{red}{$t^* > 1.50 \times 10^{-3}$} \\
 \hline
\end{tabular}
\caption{Bounds on $t^*$, the minimal violation of Hankel matrix positivity on the monodromy plane, at different derivative orders $k_{\text{max}}$ and spacetime dimensions $D$. Bounds highlighted in red indicate a certificate of infeasibility. At $k_{\text{max}} = 10$, dimensions greater than $D_{\text{crit}} = 10$ are excluded.}
\label{crit_dim_results}
\end{table}

\noindent\textbf{Extremal spectrum}

We now turn to the spectrum. Since at the critical dimension, the two-sided bounds hug the superstring values closely, it is interesting to study the zeros of the extremal functional for the upper and lower bounds. Note that due to the nature of the monodromy relations, we expect that the spectrum should be integer-spaced. Thus the interest would be in the spin distribution. In particular, whether the separation of leading and subleading trajectories can be found. 

\begin{figure}[h]
    \centering
    \includegraphics[width = 0.8 \linewidth]{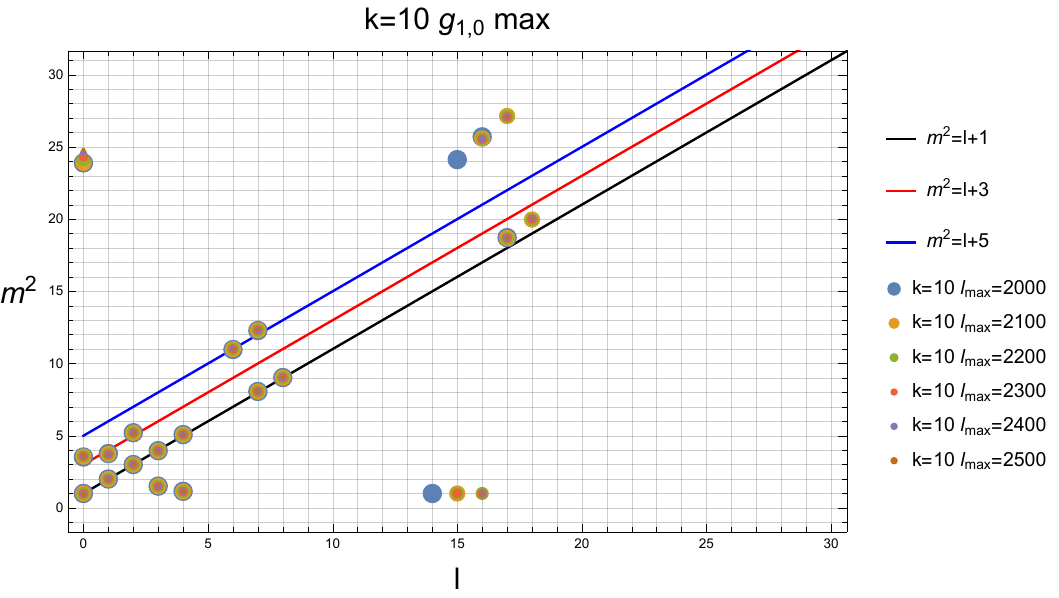}
    \caption{Extremal spectrum of max $g_{3,0}$ at $k=10$. Space-time dimension $D = 10$.}
    \label{figure: g10_max_spectrum}
\end{figure}

\begin{figure}[h]
    \centering
    \includegraphics[width = 0.8 \linewidth]{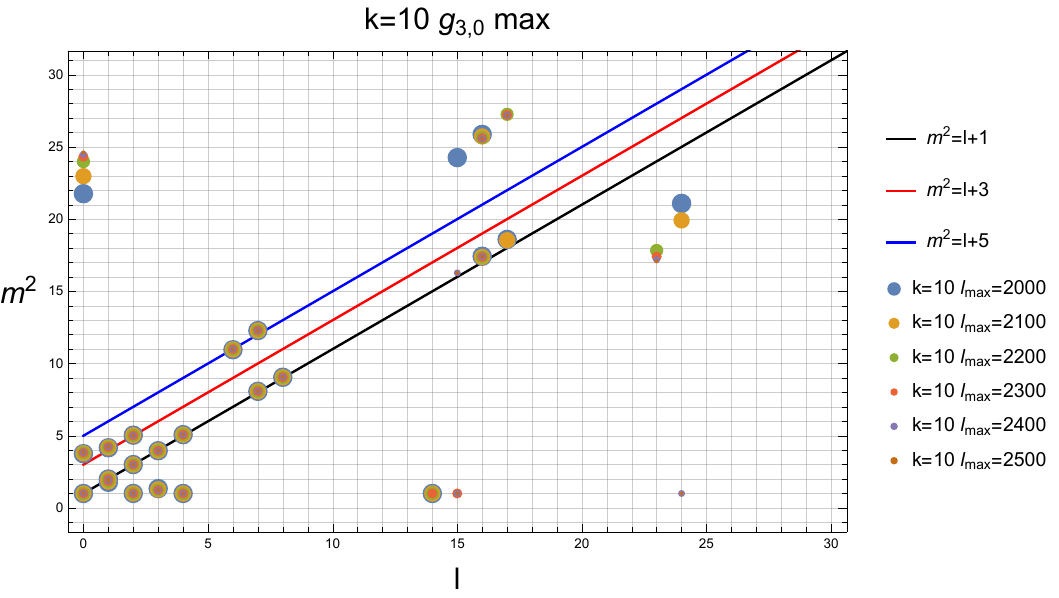}
    \caption{Extremal spectrum of min $g_{3,0}$ at $k=10$. Space-time dimension $D = 10$.}
    \label{figure: g30_min_spectrum}
\end{figure}

We analyze the extremal spectra at the upper bound of $g_{1,0}$ and the lower bound of $g_{3,0}$ since they are both very close to the superstring value, with a relative error $\left|\frac{\text{max }\tilde{g}-\tilde{g}^{\text{type-I}}}{\tilde{g}^{\text{type-I}}}\right|\lesssim O(10^{-7})$. We expect the spectra at those two bounds should reflect information about the superstring spectra. The spectra are plotted in figure. \ref{figure: g10_max_spectrum} and \ref{figure: g30_min_spectrum}

As one can see, some of the states derived from various spin truncations are stable while other states shift around. We expect that the stable states against spin truncation should reflect the spectrum of string theory. We observe that the stable states are low spin and low mass ($\ell<20$, and $m^2<30$), and these stable states can be organized into three linear trajectories, with the exception of a few states. The majority of the stable states lie on the leading trajectory while three and two states lie on the subleading and sub-subleading trajectories. The pattern does not match with the intuition that the states should fill up from low mass, since the lowest state should be the major contribution to the EFT. Instead, the states are filled in a sequence that starts from the leading trajectory to the subleading and sub-subleading trajectories. The stable states that are below the leading trajectories do not belong to the string spectrum and cannot be removed by increasing spin truncation, we anticipate that they will be ruled out by imposing higher $k$ monodromy constraints.

\section{Non-supersymmetric scalar EFT}\label{sec: None-susy}
Without supersymmetry, the low energy ordered amplitude is parameterized as:
 \begin{equation}\label{eq: ScalarEFT}
        \quad A(s, t) =  - \left( \frac{s}{t} + \frac{t}{s} \right) + b \left( \frac{1}{s} + \frac{1}{t} \right) + \sum_{k,q \geq 0} g_{k,q} s^{k-q} t^q.
    \end{equation}
Consequently, the monodromy relations now implies 
 $b=0$ and imposes the following linear relations:
\begin{align}\label{eq: NoneSusyEFT}
    &\begin{pmatrix}
        g_{0,0}\\
        g_{1,0} & g_{1,1}\\
        g_{2,0} & g_{2,1} & g_{2,2}\\
        g_{3,0} & g_{3,1} & g_{3,2} & g_{3,3}\\
    \end{pmatrix}
    = \begin{pmatrix}
        -1\\
        g_{1,0} & g_{1,0}\\
        \frac{\pi^2}{6} & \frac{\pi^2}{6} & \frac{\pi^2}{6}\\
        g_{3,0} & 2 g_{3,0} {-}\frac{\pi^2}{6} g_{1,0}  & 2g_{3,0} {-}\frac{\pi^2}{6} g_{1,0} & g_{3,0} \\
    \end{pmatrix}    
\end{align}
Note that since the dispersive representation is only valid for $g_{k,q}$ with $k{-}q\geq2$, the bound for $g_{1,0}$ can only be inferred from that of $g_{3,1}$ since $g_{3,1}=2 g_{3,0} {-}\frac{\pi^2}{6} g_{1,0} $ from eq.(\ref{eq: NoneSusyEFT}).
\subsection{EFT space and critical dimensions}
\begin{figure}[h]
    \centering
    \includegraphics[width = 0.8 \linewidth]{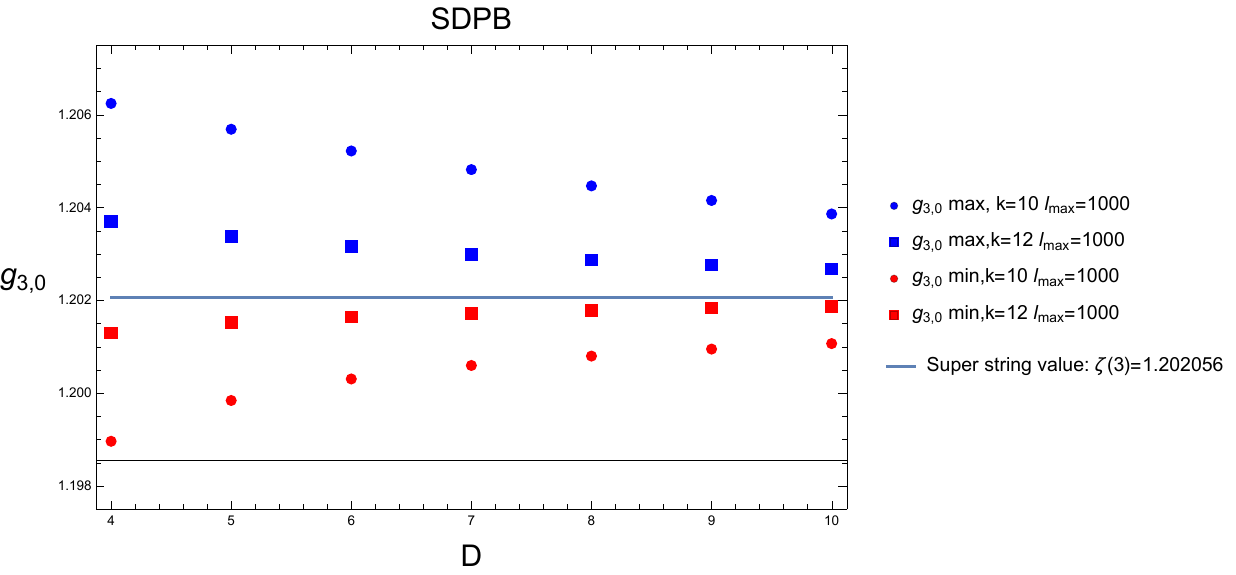}
    \caption{Upper and lower bound of $g_{3,0}$ at various space-time dimension. The results of imposing $k=10$ and $k=12$ monodromy constraints are shown in the graph. The bounds are stabilized against spin truncation.}
    \label{fig: nonsusy_g30_dimensionplot}
\end{figure}

\begin{figure}[h]
    \centering
    \includegraphics[width = 0.8 \linewidth]{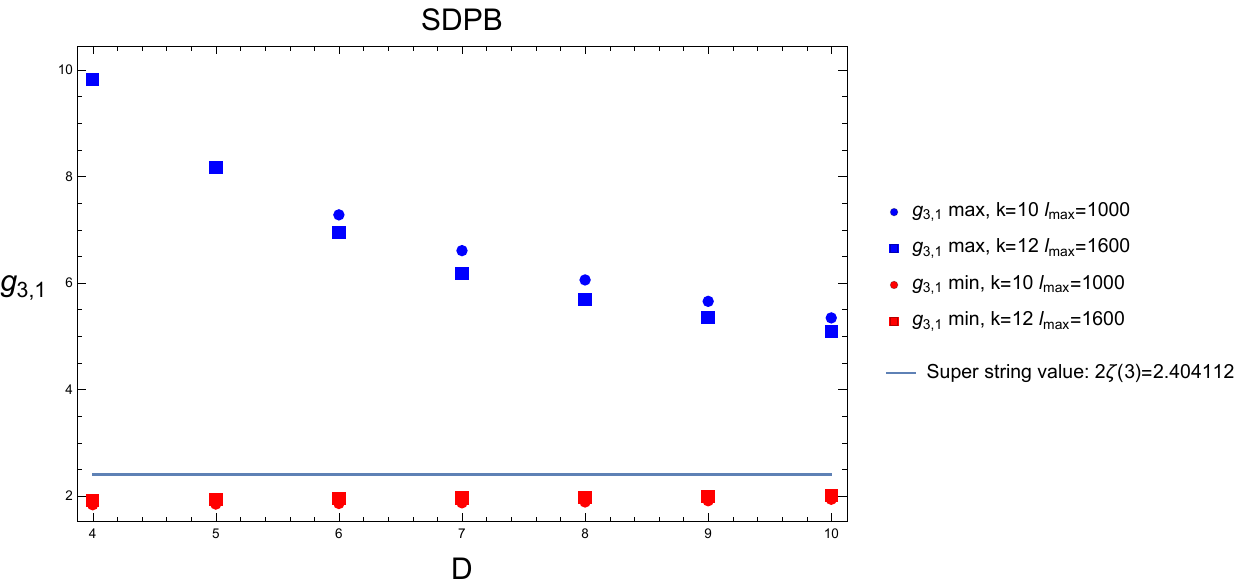}
    \caption{Upper and lower bound of $g_{3,1}$ at various space-time dimension. The results of imposing $k=10$ and $k=12$ monodromy constraints are shown in the graph. The bounds are stabilized against spin truncation.}
    \label{fig: nonsusy_g31_dimensionplot}
\end{figure}

Let us focus on $g_{3,0}$ and $g_{3,1}$, which are the two operators associated with $D^6\phi^4$. We again find tight two-sided bounds for each of the couplings that depend on the dimensions. The results are shown in fig.\ref{fig: nonsusy_g30_dimensionplot} and fig.\ref{fig: nonsusy_g31_dimensionplot}. We give their current value up to $k=12$ and $\ell_{max}=1600$ is presented in the following table 


$$
    \begin{tabular}{|c|c|c|c|c|c|}
    \hline
       D & $g_{k,q}$ & Two-sided bound & Superstring value & Relative error  \\
       \hline
       4 & $g_{3,0}$ & $1.2012 < g_{3,0} < 1.20369$ & $1.202056$ & $2.0 \times 10^{-3}$ \\
       \hline
       4 & $g_{3,1}$ & $1.91 < g_{3,1} < 9.8125$ & $2.404$ & $3.2$ \\
       \hline
       10 & $g_{3,0}$ & $1.20186 < g_{3,0} < 1.20266$ & $1.202056$ & $6.6 \times 10^{-4}$ \\
       \hline
       10 & $g_{3,1}$ & $1.99 < g_{3,1} < 5.09$ & $2.404$ & $1.2$ \\
       \hline
    \end{tabular}$$
The type-I string amplitude for massless scalars can be obtained by setting all polarization vectors orthogonal to the four momenta, i.e. $\epsilon\cdot k=0$. This leads to the four-point amplitude given by 
\begin{equation}\label{eq: SUSY_scalar_amplitude}
A(s,t)=(s^2{+}t^2{+}u^2)\frac{\Gamma[{-}s]\Gamma[{-}t]}{\Gamma[1{+}u]}\,.
\end{equation}
The bounds on $g_{3,0}$ are tight such that the relative error to string theory is $\lesssim O(10^{-3})$, however, the relative error of $g_{3,1}$ remains order 1 as we increase derivative order. Unlike the maximal SUSY case, the coefficient $g_{3,1}$ cannot be pinned down completely.

\begin{figure}[h]
    \centering
    \includegraphics[width = 0.8 \linewidth]{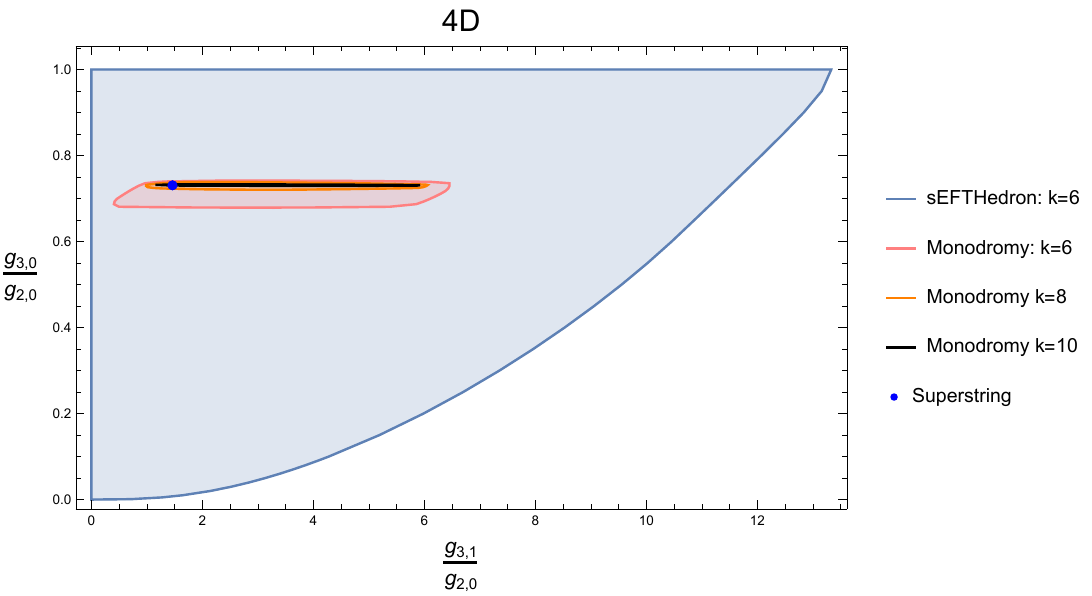}
    \caption{Comparing the allowed region of $\left(\frac{g_{3,0}}{g_{2,0}},\frac{g_{3,1}}{g_{2,0}}\right)$ by imposing crossing relations alone (sEFTHedron \cite{Chiang:2021ziz}) and imposing monodromy relations (up to $k{=}10$). The bounds on coupling $g_{3,0}$ shrink toward superstring value $\zeta(3)$, however, the bounds on $g_{3,1}$ remain finite.}
    \label{fig: nonsusy_g30g31_plot}
\end{figure}

We can further plot out the two-dimensional space $(g_{3,1},g_{3,0})$ of 
the open string amplitude. The coefficient $g_{2,0}$ is completely fixed by the monodromy relation to $\frac{\pi^2}{6}$, therefore it is equivalent to plot out the space $(\frac{g_{3,1}}{g_{2,0}},\frac{g_{3,0}}{g_{2,0}})$. We compare the result with the forward limit unitarity bounds derived in~\cite{Chiang:2021ziz} in fig.\ref{fig: nonsusy_g30g31_plot}. Note that the shape of the allowed region of open string Wilson coefficients is an extremely thin line. As we will see in the next subsection, by allowing tachyon states one can deviate significantly from the thin line in fig.\ref{fig: nonsusy_g30g31_plot}.

Finally, following the superstring analysis, we consider the critical dimension. Again, by restricting the Wilson coefficients on the monodromy plane, we minimize the violation of positivity constraints to determine if a spacetime dimension is feasible. The current best result is 
\begin{equation}
    D \leq 12, 
\end{equation}
achieved at \(k_{\text{max}} = 26\), shown in Table \ref{crit_dim_non_SUSY_results}. These results are robust, and optimal up to a relative primal-dual gap of at least
\begin{equation}
    \left| \frac{\text{primal obj \,-\, dual obj}}{\text{primal obj}} \right| < 10^{-6}, 
\end{equation}
which is treated as an indication of convergence. 

In our case, even with redundant variables explicitly eliminated (so that the Wilson coefficients are exactly placed on the monodromy plane), the computation time of the SDP roughly doubles as we increase the derivative order $k_{\text{max}}$ by 2. To rule out even lower spacetime dimensions is a work in progress, and it will be very interesting to see if the critical dimension is again set to $10$ as in the superstring case.

\begin{table}[h]
\center
\begin{tabular}{|l|l|l|l|l|}
 \hline
 & $D = 12$ & $D = 13$ & $D = 14$ & $D = 15$ 
 \\
 \hline
 $k_{\text{max}} = 20$ & $\cdots$ & $\cdots$ & \color{red}{$t^* = 2.81 \times 10^{-15}$} & \color{red}{$t^* = 7.37 \times 10^{-10}$}
 \\
 \hline
 $k_{\text{max}} = 22$ & $\cdots$ & $t^* = -7.62 \times 10^{-10}$ & \color{red}{$t^* = 1.64 \times 10^{-10}$} & $\cdots$ 
 \\
 \hline
 $k_{\text{max}} = 26$ & $t^* = -8.998 \times 10^{-25}$ & \color{red}{$t^* = 2.276 \times 10^{-14}$} & $\cdots$ & $\cdots$
 \\
 \hline
\end{tabular}
\caption{Minimal violation of unitarity in various spacetime dimensions. The current best bound is $D \leq 12$, obtained at derivative order $k_{\text{max}} = 26$. }
\label{crit_dim_non_SUSY_results}
\end{table}


\subsection{EFT with Tachyons }
\begin{figure}[h]
    \centering
    \includegraphics[width = 0.8 \linewidth]{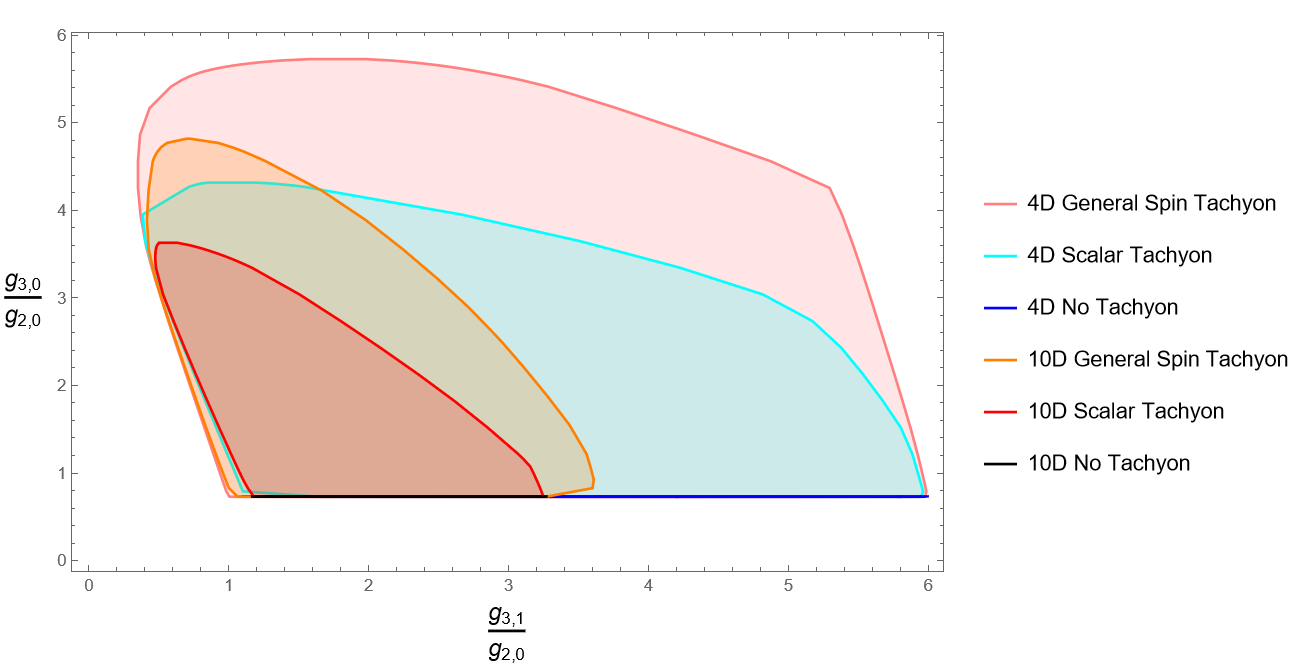}
    \caption{Comparing the allowed region of $\left(\frac{g_{3,0}}{g_{2,0}},\frac{g_{3,1}}{g_{2,0}}\right)$, by assuming different spin configuration of the tachyon state and different space-time dimension.}
    \label{fig: tachyon_nonsusy_g30g31_plot}
\end{figure}

While for physical setups we don't expect Tachyons, in anticipation of obtaining effective field theory for heterotic string, we will consider the case where there is a Tachyon in the open string spectrum. We proceed by modifying the dispersion relations to include the presence of Tachyon, namely 
\begin{equation}
A(s,t)=-\sum_{\ell}\int dm^2\;\left( \delta(m^2+M^2)\rho_{\ell}\frac{G^D_{\ell}(\theta)}{s-m^2}{+}\rho_{\ell}(m^2)\frac{G^D_{\ell}(\theta)}{s-m^2}\right)
\end{equation}
where we assume $\rho_{\ell}(m^2)\neq0$ for $m^2\geq M^2$.  Note that we considered two scenarios, the tachyon state being arbitrary spin and restricted to being scalar. The scalar tachyon bound is enclosed by the arbitrary spin tachyon bound which is consistent. We summarize the results for the couplings $g_{3,0},\, g_{3,1}$ from unitarity at $D=4,10$. Monodromy relations up to $k=10$ are used and the spin truncation is $\ell_{max}=1100$
$$\begin{tabular}{|c|c|c|c|c|c|}
    \hline
       D & $g_{k,q}$ & Arbitrary spin tachyon bound & Bosonic string value & Relative error  \\
       \hline
       4 & $g_{3,0}$ & $1.19236 < g_{3,0} < 9.429$ & $3.20205$ & $2.5$ \\
       \hline
       4 & $g_{3,1}$ & $0.57426 < g_{3,1} < 9.859$ & $1.46931$ & $6.3$ \\
       \hline
       10 & $g_{3,0}$ & $1.19893 < g_{3,0} < 7.926$ & $3.20205$ & $2.1$ \\
       \hline
       10 & $g_{3,1}$ & $0.683 < g_{3,1} < 5.947$ & $1.46391$ & $3.5$ \\
       \hline
       D & $g_{k,q}$ & Scalar tachyon bound & Bosonic string value & Relative error  \\
       \hline
       4 & $g_{3,0}$ & $1.1989 < g_{3,0} < 7.11$ & $ 3.20205$ & $1.8$ \\
       \hline
       4 & $g_{3,1}$ & $0.630 < g_{3,1} < 9.8262$ & $1.46931$ & $6.2$ \\
       \hline
       10 & $g_{3,0}$ & $1.20106 < g_{3,0} < 5.988$ & $3.20205$ & $1.5$ \\
       \hline
       10 & $g_{3,1}$ & $0.78 < g_{3,1} < 5.344$ & $1.46391$ & $3.1$ \\
       \hline
    \end{tabular}$$
In the above, we compare the bound and the bosonic string amplitude obtained from 
\begin{equation}\label{eq: Bosonic_scalar_amplitude}
    \frac{\Gamma ({-}s) \Gamma ({-}t)}{\Gamma ({-}s{-}t{+}1)} \left(\frac{tu}{s{+}1}+\frac{su}{t{+}1}+\frac{s t}{u{+}1}\right).
\end{equation}
Once again this corresponds to the Yang-Mills amplitude for bosonic string with all polarization vectors set perpendicular to the external momenta. Note that even restricting to scalar Tachyons, the boundaries are of $\mathcal{O}(1)$ deviation from the known bosonic string value. Since $g_{2,0}$ is fixed via monodromy, we can plot the 2D region for $(\frac{g_{3,1}}{g_{2,0}},\frac{g_{3,0}}{g_{2,0}})$ in fig.(\ref{fig: tachyon_nonsusy_g30g31_plot}). As one can see, substantial deviations in $\frac{g_{3,0}}{g_{2,0}}$ can be obtained via the inclusion of tachyon states.

\section{Four-dimensional gluon EFT}\label{sec: Vectors}
Let us now consider the gluon EFT. Again the disk integral would be of the form 
\begin{equation}
A(p_i, \epsilon_i)=\int_{z_1}^{z_{3}} dz_2 \frac{1}{J} \prod_{i\neq j} |z_{ij}|^{-k_{ij}} f^{\rm disk}(z_i, \bar{z}_i, k_i, \epsilon_i)
\end{equation}
Let's again assume that $f^{\rm disk}(z_i, \bar{z}_i, k_i, \epsilon_i)$ is permutation invariant. We will be interested in four dimensions, so we choose the helicity for the external states as $(1^-,2^-,3^+,4^+)$. We then immediately have 
\begin{eqnarray}
A(1^-3^+2^-4^+)={-}\mathbf{Re}\left[e^{i \pi (k_2\cdot k_3) }A(1^-2^-3^+4^+){+}e^{i \pi ({-}k_2\cdot k_4)}A(2^-1^-3^+4^+)\right] \nonumber\\
0=\mathbf{Im}\left[e^{i \pi (k_2\cdot k_3) }A(1^-2^-3^+4^+){+}e^{i \pi ({-}k_2\cdot k_4)}A(2^-1^-3^+4^+)\right] 
\end{eqnarray}
The monodromy relations now involve two sets of amplitudes, that are independent under relabeling: 
\begin{eqnarray}
A(1^-2^+3^-4^+)&=&\langle13\rangle^2[24]^2\left(\frac{-1}{st}{+}\sum_{k,q}b_{k,q}s^{k{-}q}t^q \right)\nonumber\\
A(1^-2^-3^+4^+)&=&\langle12\rangle^2[34]^2\left(\frac{-1}{st}{+}\frac{\tilde{a}}{s}{+}\tilde{c}\frac{t}{s}{+}\sum_{k,q}\tilde{b}_{k,q}s^{k{-}q}t^q \right)\,.
\end{eqnarray}
$(\tilde{a},\tilde{c})$ represents contributions from $F^2\phi$ and $F^3$ interactions.

The monodromy relation implies that $ \Tilde{a} {=} 0$ and the following linear relation between coefficients. 
\begin{align}\label{eq: VectorMono}
    &\begin{pmatrix}
        b_{0,0}\\
        b_{1,0} & b_{1,1}\\
        b_{2,0} & b_{2,1} & b_{2,2} \\
        b_{3,0} & b_{3,1} & b_{3,2} & b_{3,3} \\
    \end{pmatrix}{=}    \begin{pmatrix}
        {-} \Tilde{c} {+} \frac{\pi^2}{6}\\
        \Tilde{b}_{1,0} & \Tilde{b}_{1,0}\\
        \frac{\pi^4} {45} {-} \Tilde{b}_{2,0} & \frac{\Tilde{c} \pi^2}{6} {+} \frac{\pi^4} {40} {-} 2 \Tilde{b}_{2,0} & \frac{\pi^4} {45} {-} \Tilde{b}_{2,0}\\
        \Tilde{b}_{3,0} & {-}\frac{\pi^2}{4}\Tilde{b}_{1,0} {+} 3\Tilde{b}_{3,0} {-} \frac{1}{2}\Tilde{b}_{3,2} & {-}\frac{\pi^2}{4}\Tilde{b}_{1,0} {+} 3\Tilde{b}_{3,0} {-} \frac{1}{2}\Tilde{b}_{3,2} & \Tilde{b}_{3,0}\\
    \end{pmatrix}\nonumber\\
    &\begin{pmatrix}
        \Tilde{b}_{0,0}\\
        \Tilde{b}_{1,0} & \Tilde{b}_{1,1}\\
        \Tilde{b}_{2,0} & \Tilde{b}_{2,1} & \Tilde{b}_{2,2}\\
        \Tilde{b}_{3,0} & \Tilde{b}_{3,1} & \Tilde{b}_{3,2} & \Tilde{b}_{3,3}\\
    \end{pmatrix} {=}    \begin{pmatrix}
        \Tilde{c} {+} \frac{\pi^2}{6}\\
        \Tilde{b}_{1,0} & \Tilde{b}_{1,0}\\
        \Tilde{b}_{2,0} & {-} \frac{\Tilde{c} \pi^2}{6} {-} \frac{\pi^4} {120} {+} \Tilde{b}_{2,0} & {-}\frac{\Tilde{c} \pi^2}{6} {+} \frac{\pi^4}{90}\\
        \Tilde{b}_{3,0} & {-}\frac{\pi^2}{12} \Tilde{b}_{1,0}{+} \Tilde{b}_{3,0} {+} \frac{1}{2} \Tilde{b}_{3,2} & \Tilde{b}_{3,2} & \frac{\pi^2}{12}\Tilde{b}_{1,0} {+} \frac{1}{2} \Tilde{b}_{3,2}\\
    \end{pmatrix}\nonumber\\
\end{align}

In general, for particles carrying external spins, partial wave expansion is more involved as the irreps can now involve mixed representations. Recall that in scattering scalars the only vector in a given channel that can be used to span irreps is the difference of external momenta on one side of the channel. Thus one can only allow for spin-$\ell$ irreps. For the scattering of vectors, the polarization vectors provide an additional basis, and thus mixed representations are allowed (see~\cite{Caron-Huot:2022jli}). However, in four-dimensions things simplify and once again only spin-$\ell$ irreps are allowed~\cite{Arkani-Hamed:2022gsa}, with the partial waves expanded on the Wigner $d$-matrices that depend on the external helicity. The two amplitude has the following dispersion relations:
\begin{eqnarray}
A(1^-2^+3^-4^+)&=&{-}u^2\sum_l\int dm^2\frac{d^{\ell\geq2}_{-2,-2}(\theta)}{\cos^4\frac{\theta}{2}m^4}\frac{\rho^{+-}_{\ell}(m^2)}{s-m^2}\nonumber\\
A(1^-2^-3^+4^+)&=&{-}s^2\sum_l\int dm^2\frac{d^{\ell\geq0}_{0,0}(\theta)}{m^4}\frac{\rho^{++}_{\ell}(m^2)}{s-m^2}
\end{eqnarray}
The two dispersion relations are simply the interchange of $s$ and $u$-channel contributions. The above is modified for a Tachyon as follows 
\begin{align}
A(1^-2^+3^-4^+)&={-}u^2\sum_l\int dm^2\frac{d^{\ell\geq2}_{-2,-2}(\theta)}{\cos^4\frac{\theta}{2}m^4}\frac{|p_{\ell}(m^2)^{+-}|^2}{s-m^2}\nonumber\\
A(1^-2^-3^+4^+)&={-}s^2\sum_l\int dm^2\left(\delta(m^2{+}M^2)\frac{d^{\ell\geq0}_{0,0}(\theta)}{m^4}\frac{|p_{\ell}^{++}|^2}{s-m^2}+\frac{d^{\ell\geq0}_{0,0}(\theta)}{m^4}\frac{|p_{\ell}(m^2)^{++}|^2}{s-m^2}\right)
\end{align}
\subsection{Gluon EFT}
We consider bounds on the couplings $\tilde{c}, \tilde{b}_{1,0},\, \tilde{b}_{2,0}$ ,and $\tilde{b}_{3,2}$ representing operators $F^3$, $D^2F^4$, $D^4F^4$ and $D^6F^4$ respectively. Note that the coefficient for $F^4$, i.e. $\tilde{b}_{1,0}$ are determined by $\tilde{c}$ through monodromy relations as seen from eq.(\ref{eq: VectorMono}). The results from unitarity are summarized in the following table. Monodromy relations up to $k=10$ are used and the spin truncation is $\ell_{max}=1500$.

$$\begin{tabular}{|c|c|c|c|}
            \hline
            Wilson coefficients & Two-sided bound & Superstring value & Relative error\\\hline
            $\tilde{b}_{1,0}$ & $1.20203<\tilde{b}_{1,0}<1.202059$ & $1.202056$ & $2.4{\times}10^{-5}$\\\hline
            $\tilde{b}_{2,0}$ & $1.08231<\tilde{b}_{2,0}<1.08233$ & $1.082323$ & $1.8{\times}10^{-5}$\\\hline
            $\tilde{b}_{3,2}$ & $0.09653<\tilde{b}_{3,2}<0.0966$ & $0.09655$ & $7.2{\times}10^{-4}$\\\hline
            $\tilde{c}$ & $-1{\times}10^{-6}<\tilde{c}<2{\times}10^{-5}$ & $0$ & N/A\\\hline
        \end{tabular}$$
As one can see, remarkably, \textit{the Wilson coefficients of gluon EFT are pinned down to the superstring values even without assuming SUSY}! In particular, the $F^3$ operator is set to zero similar to $\phi F^2$. Thus we see that for vectors, monodromy relations imply non-trivial relations amongst distinct helicity ordered amplitude, and by requiring unitarity pins down the superstring value. To deviate from superstring while maintaining unitarity, one has to consider set up with Tachyonic states or closed string theory, i.e. the heterotic string. 
\subsection{Gluon EFT with Tachyons}
Similarly, we can also consider gluon EFT with tachyon exchange. The bounds on the couplings $\tilde{b}_{1,0},\, \tilde{b}_{2,0},\, \tilde{b}_{3,2},\, \text{and } \tilde{c}$ from unitarity. Monodromy relations up to $k=10$ are used and the spin truncation is $\ell_{max}=1500$.
$$
        \begin{tabular}{|c|c|c|c|}
            \hline
            Wilson coefficients & Arbitrary spin tachyon bound & Bosonic string value &  Relative error \\\hline
            $\tilde{b}_{1,0}$ & $1.20193<\tilde{b}_{1,0}<2.208$ & $2.20205$ & $0.45$\\\hline
            $\tilde{b}_{2,0}$ & $0.07817<\tilde{b}_{2,0}<1.0827$ & $0.08232$ & $12.2$\\\hline
            $\tilde{b}_{3,2}$ & $0.0959<\tilde{b}_{3,2}<0.8592$ & $0.85573$ & $7.9$\\\hline
            $\tilde{c}$ & $-1.00418<\tilde{c}<3.3{\times}10^{-4}$ & $-1$ & $1.0$\\\hline
            Wilson coefficients & Scalar tachyon bound & Bosonic string value & Relative error\\\hline
            $\tilde{b}_{1,0}$ & $1.20203<\tilde{b}_{1,0}<2.204$ & $2.20205$ & $0.45$\\\hline
            $\tilde{b}_{2,0}$ & $0.0802<\tilde{b}_{2,0}<1.08233$ & $0.08232$ & $12.2$\\\hline
            $\tilde{b}_{3,2}$ & $0.0964<\tilde{b}_{3,2}<0.8573$ & $0.85573$ & $7.9$\\\hline
            $\tilde{c}$ & $-1.002<\tilde{c}<2{\times}10^{-5}$ & $-1$ & $1.0$\\\hline
        \end{tabular}$$
        
\begin{figure}[h]
    \centering
    \includegraphics[width = 0.7 \linewidth]{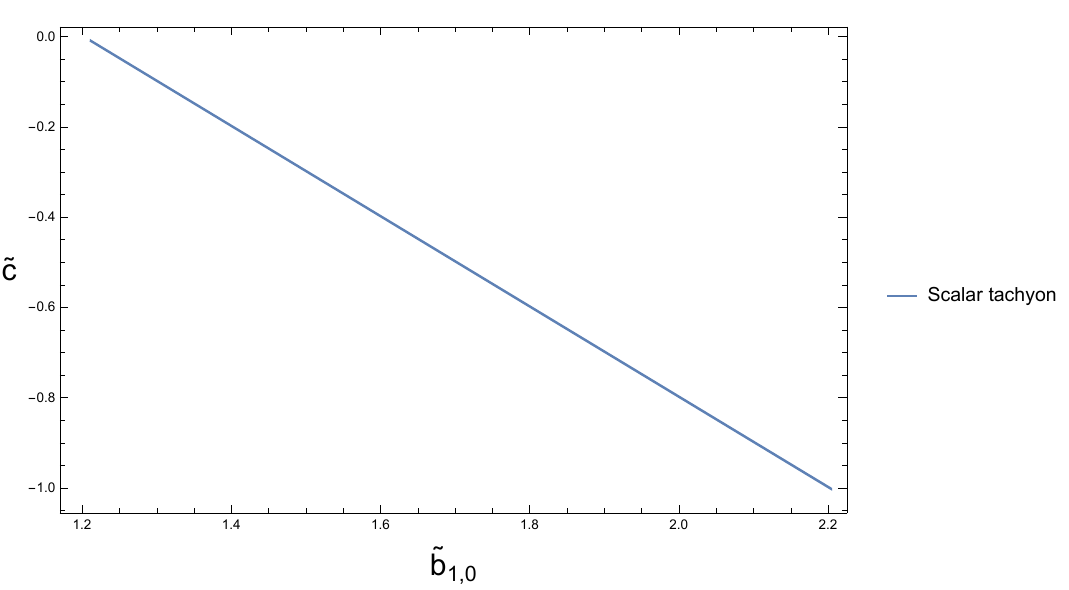}
    \caption{The allowed region of $\left(\tilde{b}_{1,0},\tilde{c}\right)$. This is a two-dimensional region but the width of the region is small. For most fixed $\tilde{b}_{1,0}$, the discrepancy between of the bound of $\tilde{c}$ is $\left|\frac{\tilde{c}_{max}-\tilde{c}_{min}}{\tilde{c}_{max}}\right|\sim O(10^{-3}
    )$.}
    \label{fig: scalar_tachyon_b10c_plot}
\end{figure}

\begin{figure}[h]
    \centering
    \includegraphics[width = 0.7 \linewidth]{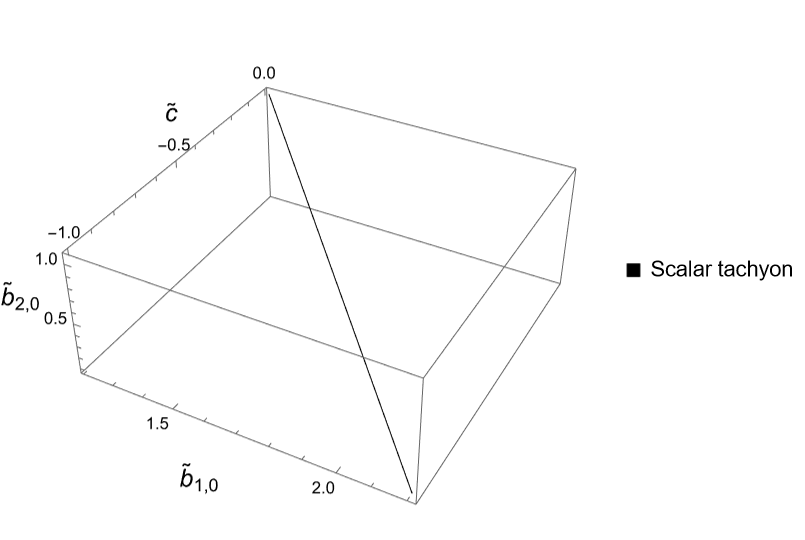}
    \caption{The allowed region of $\left(\tilde{b}_{1,0},\tilde{c},\tilde{b}_{2,0}\right)$. This is a three-dimensional region but the width of the region is small. For most fixed $\tilde{b}_{1,0}$, the discrepancy between of the bound of $\tilde{b}_{2,0}$ is $\left|\frac{\tilde{b}_{2,0}^{\text{max}}-\tilde{b}_{2,0}^{\text{min}}}{\tilde{b}_{2,0}^{\text{min}}}\right|\sim O(10^{-3}
    )$.}
    \label{fig: scalar_tachyon_b10cb20_plot}
\end{figure}

Note that for the scalar tachyon bootstrap, the bosonic string value is close to one end of the bounds while the superstring value is close to the other, this suggests that the allowed space of the scalar tachyon bootstrap could be spanned by superstring and bosonic string. We summarize the bounds and string values in table \ref{table: gluon_scalar_tachyon_bound_compare_with_superstring_bosonicstring}. However, since there are infinitely many Wilson coefficients, the fact that the one-dimensional projection of the entire space is mostly spanned by bosonic and superstring theory could be just a coincidence. We further carve out the two and three-dimensional space to compare the allowed region and the span of bosonic string and superstring. The plots are shown in fig. \ref{fig: scalar_tachyon_b10c_plot} and \ref{fig: scalar_tachyon_b10cb20_plot}. The allowed region in the two and three-dimensional space is a very narrow line, supporting the conjecture that the space is spanned by bosonic string and superstring amplitudes.
\begin{table}[h]
\centering
        \begin{tabular}{|c|c|c|}
            \hline
            Wilson coefficients & Error of bosonic string & Error of Superstring \\\hline
            $\tilde{b}_{1,0}$ &  (Max)\;$8.8\times10^{-4}$ & (Min)\;$1.6\times10^{-5}$\\\hline
            $\tilde{b}_{2,0}$ & (Min)\;$2.5\times10^{-2}$ & (Max)\;$9.2\times10^{-6}$ \\\hline
            $\tilde{b}_{3,2}$ &(Max)\;$1.8\times10^{-3}$ & (Min)\;$1.5\times10^{-3}$ \\\hline
            $\tilde{c}$ & (Min)\;$2\times10^{-3}$ &(Max)\;N/A \\\hline
        \end{tabular}
        \caption{The relative error of scalar tachyon bounds and string theory. The relative error of $\tilde{c}$ at superstring is unavailable since $\tilde{c}^{\text{ss}}=0$. The relative error is defined by $\left|\frac{\tilde{b}^{\text{bound}}_{k,q}-\tilde{b}^{\text{ss/bs}}_{k,q}}{\tilde{b}^{\text{ss/bs}}_{k,q}}\right|$.}
        \label{table: gluon_scalar_tachyon_bound_compare_with_superstring_bosonicstring}
\end{table}

\section{Gravitational EFT}\label{sec: GraviEFT}
Equipped with the open string EFT, we are now ready to uplift this to their gravitational counterpart. Since we assume trivial monodromy, we can straightforwardly utilize the KLT relations eq.(\ref{eq: KLT}). Note that the result will be automatically crossing symmetric since  $A(s,t)$ solves the monodromy relations. Indeed 
\begin{equation}
M(s,t)=8G\sin(\pi t)A(s,t)A(t,u)=8G\sin(\pi u)A(s,u)A(t,u)=M(s,u)\,.
\end{equation}
However, unitarity is not guaranteed. Thus in this section, we explicitly test the compatibility of KLT double copy with unitarity.

\noindent\textbf{Four-scalar amplitude}

Let's begin with the four-scalar amplitude, where the low-energy amplitude is parameterized as 
\begin{eqnarray}
M(s,t)=8\pi G\left(\frac{tu}{s}{+}\frac{su}{t}{+}\frac{st}{u}\right){+}\lambda_1\left(\frac{1}{s}{+}\frac{1}{t}{+}\frac{1}{u}\right){+}\lambda_2{+}g_2\sigma_2{+}g_3\sigma_3{+}g_4\sigma_2^2{+}\cdots
\end{eqnarray}
where $\sigma_2=(s^2{+}t^2{+}u^2)$ and $\sigma_3=stu$. We first consider double copy none-supersymmetric scalar amplitudes. Now using eq.(\ref{eq: ScalarEFT}), with the couplings living on the monodromy plane in eq.(\ref{eq: NoneSusyEFT}), the KLT relations  eq.(\ref{eq: KLT}) give $\lambda_{1,2}=0$ and
\begin{eqnarray}
g_2=8G \pi g_{1,0}, \quad g_3=8G \pi g^2_{1,0},\quad g_4=4 G \pi g_{3,0}
\end{eqnarray}
Note that while we do not have dispersive representation for $g_{1,0}$, one can obtain bounds on $g_{1,0}$ through the linear relation $\frac{6}{\pi^2}(2g_{3,0}-g_{3,1})=g_{1,0}$ given by eq.(\ref{eq: NoneSusyEFT}). The inferred bounds are 
\begin{equation}\label{eq: g10table}
\begin{tabular}{|c|c|c|c|}
\hline
    D & No tachyon bound & Scalar tachyon bound & General tachyon bound\\\hline
    5 & $-3.5071<g_{1,0}<0.337$ & $-3.5071<g_{1,0}<7.445$ & $-3.5182<g_{1,0}<9.918$\\\hline
    6 & $-2.963<g_{1,0}<0.3307$ & $-2.963<g_{1,0}<7.241$ & $-3.001<g_{1,0}<9.622$\\\hline
    10 & $-1.786<g_{1,0}<0.28$ & $-1.786<g_{1,0}<6.757$ & $-2.077<g_{1,0}<8.989$ \\\hline
\end{tabular}\,.
\end{equation}

\begin{figure}[h]
    \centering
    \begin{subfigure}{0.49\textwidth}
         \includegraphics[width = \textwidth]{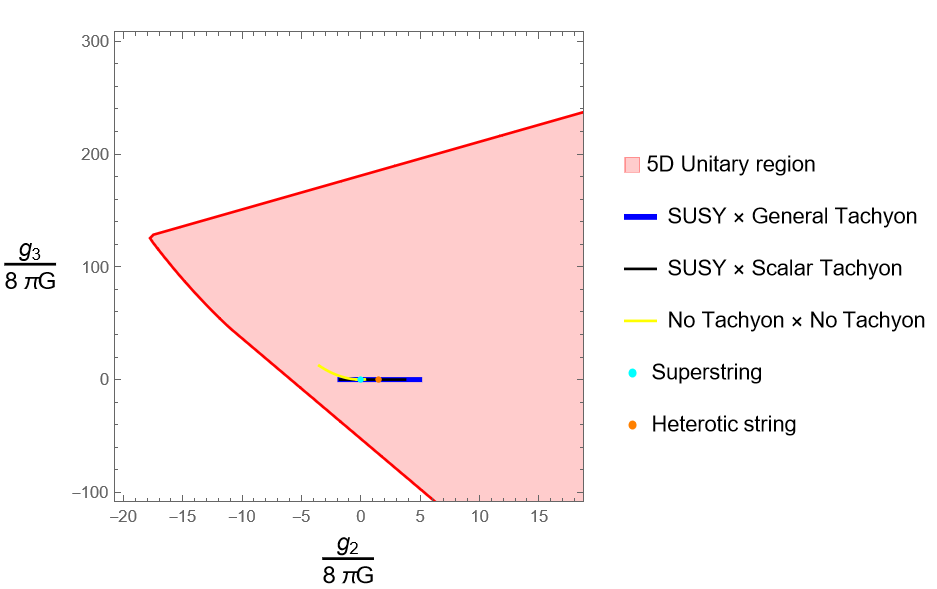}
         \captionsetup{width=.8\linewidth}
         \caption{5D unitary region and the three scenarios of KLT double copy.}
    \end{subfigure}
    \begin{subfigure}{0.49\textwidth}
         \includegraphics[width = \textwidth]{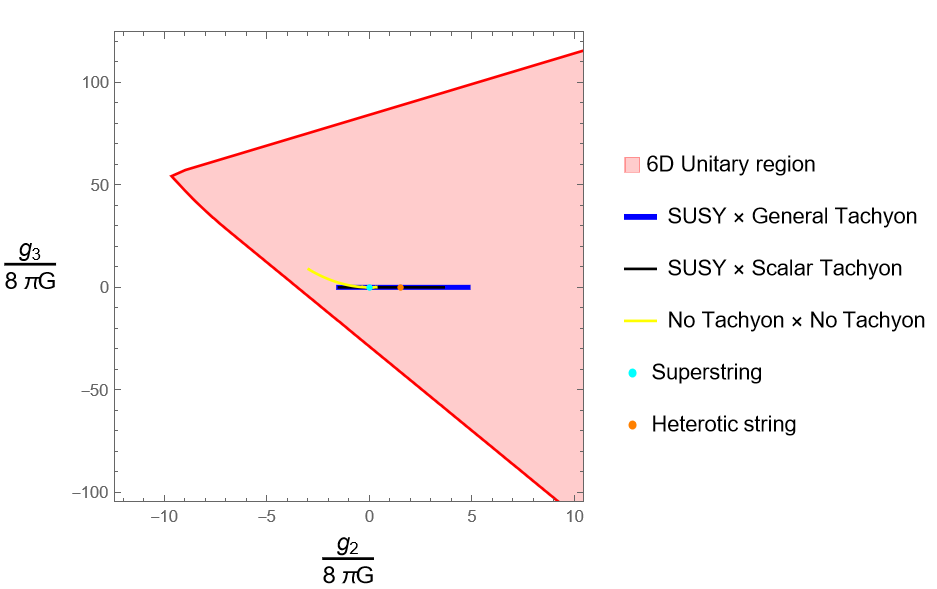}
         \captionsetup{width=.8\linewidth}
         \caption{6D unitary region and the three scenarios of KLT double copy.}
    \end{subfigure}
    \begin{subfigure}{0.5\textwidth}
         \includegraphics[width = \textwidth]{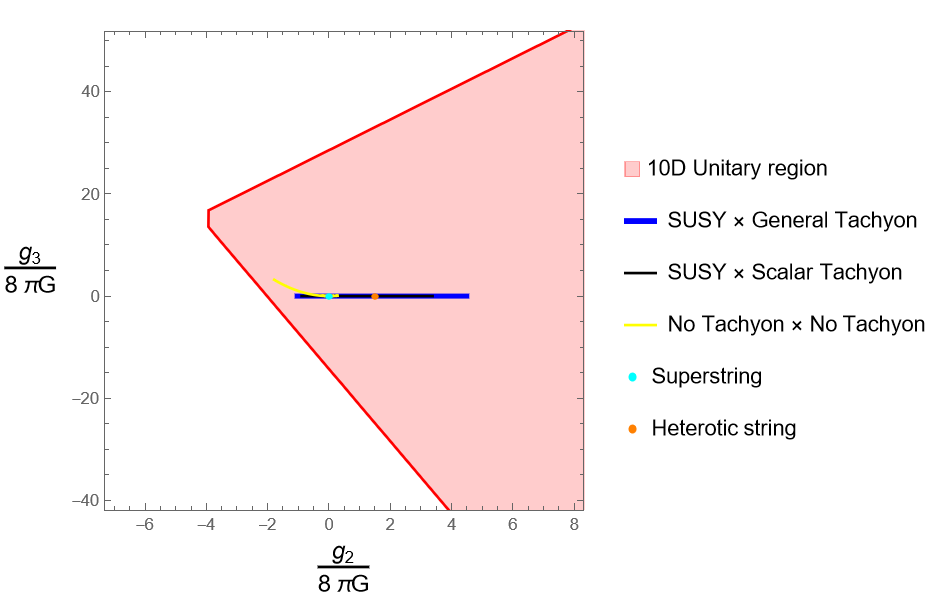}
         \captionsetup{width=.8\linewidth}
         \caption{10D unitary region and the three scenarios of KLT double copy.}
    \end{subfigure}
    \caption{Comparing the unitary space with the closed string region via KLT double copy. The unitary region of the gravitational scalar amplitude in the plot is unbounded from the right. The heterotic string value is obtained by the double copy of eqns.(\ref{eq: SUSY_scalar_amplitude}) and (\ref{eq: Bosonic_scalar_amplitude})}
    \label{fig: simon_scalar_unitary}
\end{figure}

We now map the space of open string coefficients $g_{k,q}$ to the space of closed string $(g_{2},g_{3})$. The space $(g_2,g_3)$ is thus a parabola parameterized by the coefficient $g_{1,0}$ ($D^2\phi^4$) of scalar amplitude. We compared with the unitary region on the gravitational scalar amplitude~\cite{Caron-Huot:2021rmr} in figure \ref{fig: simon_scalar_unitary}. The closed string regions are within the unitary region of the gravitational scalar amplitude. Thus unitarity of the open string couplings appears to guarantee that of the closed string.

In the above, we've also considered the heterotic construction, that is, the double copy of SUSY amplitude $A^{\text{SUSY}}$ and non-SUSY amplitude that contains tachyonic exchange $A^{\text{Tach}}$, i.e.
\begin{equation}
    M(s,t)=8G\sin(\pi t)A^{\text{SUSY}}(s,t)A^{\text{Tach}}(t,u)\,.
\end{equation}
We again have $\lambda_{1,2}=0$ and
\begin{equation}
    g_2=4G\pi g^{\text{Tach}}_{1,0},\qquad g_{3}=0,\qquad g_{4}=2G\pi (g^{\text{Tach}}_{3,0}+g^{\text{SUSY}}_{1,0})\,.
\end{equation}
The resulting $(g_2,g_3)$ space would be a horizontal line on the $g_3=0$ axis.

\noindent\textbf{Four-graviton amplitude}

For the four-graviton EFT, consider 
\begin{equation}
M(1^+,2^+,3^-,4^-)=\pi[12]^4\langle 34\rangle^4\left[\frac{1}{s t u}{+}\gamma_1\frac{1}{s}{+}\gamma_2\frac{t^2}{s}{+}\alpha_0{+}\alpha_1 s+\alpha_2t\right]
\end{equation}
where $\gamma_1$ $\gamma_2$ encodes the presence of $R^2\phi$ and $R^3$ operators separately. Using the KLT relations 
\begin{equation}
M(1^+,2^+,3^-,4^-)=sin(\pi t) A(1^+,2^+,3^-,4^-)A(1^+,3^-,2^+,4^-)\,,
\end{equation}
and monodromy relations for the photon amplitude eq.(\ref{eq: VectorMono}), we find 
\begin{equation}
\gamma_1=2\tilde{c}, \quad \gamma_2=-\tilde{c}^2,\quad \alpha_0=2\tilde{b}_{1,0}, \quad \alpha_1=2\tilde{b}_{2,0}{-}\frac{\pi^4}{45},\quad \alpha_2=-\tilde{c}^2.
\end{equation}

We can also consider the construction of the heterotic string, that is the double copy of scalar tachyon amplitude $A^{\text{s}}$ and no tachyon amplitude $A^{\text{n}}$,
\begin{equation}
    M(1^+,2^+,3^-,4^-)=\sin(\pi t)A^{\text{s}}(1^+,2^+,3^-,4^-)A^{\text{n}}(1^+,3^-,2^+,4^-).
\end{equation}
We obtain the Wilson coefficients
\begin{equation}
    \gamma_1=\tilde{c}^{\text{s}}+\tilde{c}^{\text{n}},\quad\gamma_2=- \tilde{c}^{\text{s}}\tilde{c}^{\text{n}},\quad\alpha_0=\tilde{b}_{1,0}^{\text{s}}+\tilde{b}_{1,0}^{\text{n}},\quad\alpha_1=\tilde{b}_{2,0}^{\text{s}}+\tilde{b}_{2,0}^{\text{n}}-\frac{\pi^4}{45},\quad\alpha_2=-\tilde{c}^{\text{s}}\tilde{c}^{\text{n}}.
\end{equation}
Here we denote the coefficients of the scalar tachyon amplitude and no tachyon amplitude by an over script $\tilde{b}^{\text{s/n}}$.

This time, we consider the projective space of the closed string amplitude $\left(\frac{\alpha_1}{\alpha_0},\frac{\alpha_2}{\alpha_0}\right)$ and compare it with the unitary space of the four graviton amplitude given in \cite{Chiang:2022jep}. The region is one the line $\alpha_2=0$ since the no tachyon gluon bootstrap is pinned down by superstring, supersymmetry enforces the operator $R^3$ to vanish. The two ends of the region are close to superstring and heterotic string amplitudes, this is due to the fact that the scalar tachyon gluon region coefficient space we have considered is mostly spanned by superstring and bosonic string.

\begin{figure}[h]
    \centering
    \includegraphics[width = 0.8 \linewidth]{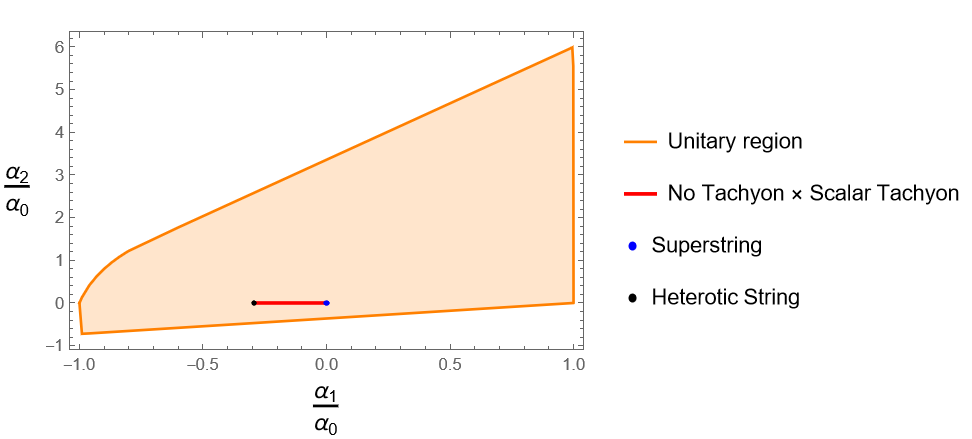}
    \caption{The allowed region of $\left(\frac{\alpha_1}{\alpha_0},\frac{\alpha_2}{\alpha_0}\right)$. The endpoints are close to superstring and heterotic string amplitude.}
    \label{fig: graviton_double_copy}
\end{figure}

\noindent\textbf{Four-photon amplitude}

For the four-photon EFT consider instead 
\begin{equation}
M(1^+,2^+,3^-,4^-)=\pi[12]^2\langle 34\rangle^2\left[\frac{s}{t u}{+}c_1\frac{1}{s}{+}\lambda \right]
\end{equation}
We can construct this amplitude using the double copy, 
\begin{equation}
\sin(\pi s) A(1^+2^+3^-4^-)A(s,u)=\pi[12]^2\langle34\rangle^2\left[\frac{s}{tu}{-}\frac{1}{s}{+}g_{1,0}\right]
\end{equation}
From the inferred bounds on $g_{1,0}$ in eq.(\ref{eq: g10table}), we see that there are no definite sign for $\lambda$ related to $F^4$ operator. This is another example of  violation of the weak gravity conjecture~\cite{Arkani-Hamed:2006emk}, which requires it to be positive. Indeed as discussed in~\cite{Caron-Huot:2021rmr, Henriksson:2022oeu}, in gravitational theories pure Einstein gravity contributes   positive time delay, and thus negative contributions from higher dimension operators are allowed from causality.  
\section{Conclusion and outlook}
In this paper, we consider the space of consistent Wilson coefficients determined from the S-matrix bootstrap, conveniently term the EFThedron, and its intersection with the monodromy subplane. The latter is the hyperplane defined by the linear relations between the Wilson coefficients that are implied by the string monodromy relations.

We argue that requiring the presence of massless poles, the monodromy for the open string is constrained to that given by the standard Koba-Nielsen factors. Given this, we demonstrate that if maximal supersymmetry is assumed, the intersection of the monodromy plane and the EFT hedron reduces to tiny islands surrounding type-I super-string EFT couplings. In particular, for the three leading coefficients, numerical SDPB methods give double-sided bounds that are within $10^{{-}4}$ of the superstring result. Importantly using the Hankel constraints, which are necessary conditions for unitarity, we are able to prove that the critical dimension is 10. This is quite remarkable given that we are only considering the EFT with a finite number of low-energy couplings. 

We also consider non-supersymmetric scalars with or without Tachyon states. In such case, while the resulting bootstrap no longer leads to islands, we obtain ``strips" if Tachyons are absent. That is, while certain EFT coefficients are confined to superstring values, others allow for finite deviations. Without assuming supersymmetry, we are able to show that the critical dimension is at most 12. 

For four-dimensional gluon external states, surprisingly we find that all couplings are again cornered to superstring values, with errors of the order $10^{-4}$. This is without assuming any supersymmetry! Deviations can only occur by introducing Tachyons in the spectrum. With scalar Tachyons, we show that the allowed region is spanned by the bosonic and superstring. 

The monodromy relations imply a double copy KLT mechanism that generates gravitation amplitudes. Using our open string EFT amplitudes, which are compatible with monodromy relations, we can straightforwardly obtain a tentative closed string EFT. While crossing symmetric is automatic, unitarity requires additional check. Thus a consistent closed string EFT is given by the intersection between the KLT image of open string EFT, and the gravitational EFThedron given in~\cite{Caron-Huot:2021rmr, Caron-Huot:2022ugt}. Interesting, so far our analysis show that unitarity for the open string automatically imply the unitary of the closed string.

Given the power of monodromy relations, an immediate question is how can we generalize? Note that the standard Koba-Nielsen factor induced monodromies are valid only if we consider the presence of massless poles in both channels of the open string scattering. This would be expected for the scattering of adjoint states. More generally we expect additional monodromies for matter scattering, where massless poles are only present in one of the channels. Indeed for matter fields arising from intersecting branes, there are new contributions to the monodromy depending on the relative angles of the branes ~\cite{Lust:2004cx}. It would be interesting to bootstrap constraints on the possible modified monodromies. 

So far we have not considered the color structure accompanying open string amplitude. Unitarity requires that the imaginary part of the amplitude must be positively expanded on the kinematic polynomials \textit{and} color projectors. Indeed recently it was shown for particular stringy inspired UV completion of colored amplitudes, such constraints lead to bounds on the gauge group~\cite{Bachu:2022gof}. For open strings, things are even more constrained due to the fact that a given amplitude is associated with a fixed color trace. For example, for the ordering $1234$ we have the following dispersion relation
\begin{equation}
tr(T^aT^bT^cT^d)A(1234)=-\int_{M^2}^\infty ds'\;\sum_{\ell, I}\frac{\rho_\ell(s')P_I^{ab;cd}G^D_{\ell}(1+2t/s')}{s-s'}
\end{equation}
where $P_I^{ab;cd}$ are the $s$-channel color projectors. There are 5 independent projectors for SO(N) and 6 for SU(N). The projectors can be converted back to the trace bases, for example for SO(N), 
\begin{equation}  
\left(\begin{array}{c} {\rm Tr}[a,b,c,d] \\ {\rm Tr}[a,b,d,c]   \\   {\rm Tr}[a,d,b,c] \\  {\rm DTr}[a,b; c,d] \\ {\rm DTr}[a,c; b,d] \\ {\rm DTr}[a,d; b,c]\end{array}\right)  =\left(\begin{array}{cccccc}\frac{1}{2} & \frac{1}{2} & \frac{1}{2} & -1 & 0 & 0 \\ \frac{N{-}1}{2} & \frac{N{-}2}{4} & 0 & 0 & \frac{N{-}2}{4} & 0 \\  \frac{N{-}1}{2} & \frac{N{-}2}{4} & 0 & 0 & \frac{2{-}N}{4} & 0 \\ 1 & 1 & 1 & 1 & 1 & 1 \\ \frac{N(N{-}1)}{2} & 0 & 0 & 0 & 0 & 0 \\ 1 & 1 & 1 & 1 & -1 & -1\end{array}\right) \left(\begin{array}{c} P^t_1 \\ P^t_2 \\ P^t_3  \\ P^t_4 \\ P^t_5  \\ P^t_6 \end{array}\right)
\end{equation}  
The absence of other single trace and all double trace introduces new ``null" constraints for the bootstrap. It will be interesting to explore these new constraints, and see if one can put bounds on the rank of the gauge group from the bottom up.

\section*{Acknowledgements}
We thank Oliver Schlotterer and Stephan Stieberger for discussions. We would like to express our gratitude to David Poland for discussions and for granting us access to the computational resources at Yale. Computations in this work were performed on the Yale Grace computing cluster, supported by the facilities and staff of the Yale University Faculty of Sciences High Performance Computing Center. We greatly appreciate Henry Liao for discussions on advising computational aspects. We thank Justin Berman, Henriette Elvang, and Aidan Herderschee for sharing their upcoming draft. LYC, YTH, and
HCW are supported by MoST Grant No. 109-2112-M-002 -020 -MY3 and 112-2811-M-002 -054 -MY2. 
\appendix




\bibliography{refs}
\bibliographystyle{JHEP}
\end{document}